\documentclass[useAMS,usenatbib]{mn2e}

\usepackage{amsmath}
\usepackage{latexsym}
\usepackage{amssymb}
\usepackage{graphicx}

\usepackage{subfigure}

\newcommand{\detg}{{\sqrt{-g}}}
\newcommand{\del}{{\partial}}

\newcommand{\msun}{{\rm M_{\odot}}}

\newcommand{\dF}{{^{^*}\!\!F}}
\newcommand{\alf}{Alfv\'en}

\newcommand{\yr}{{\rm\,yr}}

\newcommand{\cm}{{\rm\,cm}}

\def\fps@figure{bp}%
\def\fps@table{bp}%
\def\fps@plate{bp}%
\def\eps@scaling{1.0}%
\newcommand\epsscale[1]{\gdef\eps@scaling{#1}}%
\newcommand\plotone[1]{%
 \centering
 \leavevmode
 \includegraphics[width={\eps@scaling\columnwidth}]{#1}%
}%
\newcommand\plottwo[2]{%
 \centering
 \leavevmode
 \columnwidth=.45\columnwidth
 \includegraphics[width={\eps@scaling\columnwidth}]{#1}%
 \hfil
 \includegraphics[width={\eps@scaling\columnwidth}]{#2}%
}%
\newcommand\plotfiddle[7]{%
 \centering
 \leavevmode
 \vbox\@to#2{\rule{\z@}{#2}}%
 \includegraphics[%
  scale=#4,
  angle=#3,
  origin=c
 ]{#1}%
}%

\newcommand\araa{\rmfamily{ARA\&A}}%
\newcommand\apj{\rmfamily{ApJ}}%
\newcommand\apjl{\rmfamily{ApJ}}%
\newcommand\mnras{\rmfamily{MNRAS}}%
\newcommand\pasj{\rmfamily{PASJ}}%
\newcommand\nat{\rmfamily{Nature}}%
\let\la=\lesssim            
\let\ga=\gtrsim
\title[Disk-Jet Coupling in Black Hole Accretion Systems I: GRMHD Models
]{Disk-Jet Coupling in Black Hole Accretion
  Systems I: General Relativistic Magnetohydrodynamical Models}

\author[Jonathan C. McKinney and Ramesh Narayan]{Jonathan  C. McKinney$^{1}$
  \thanks{E-mail: jmckinney@cfa.harvard.edu (JCM); narayan@cfa.harvard.edu (RN)}
  and Ramesh Narayan$^{1}$\footnotemark[1]\\
  $^{1}$Institute for Theory and Computation, Center for
  Astrophysics, Harvard University, 60 Garden St., Cambridge, MA,
  02138
}

\begin{document}
  \date{Accepted 2006 November 14. Received 2006 November 13; in original form 2006 July 25}
  \pagerange{\pageref{firstpage}--\pageref{lastpage}} \pubyear{2006}
  \maketitle \label{firstpage}

  \begin{abstract}

    General relativistic numerical simulations of magnetized accretion
    flows around black holes show a disordered electromagnetic
    structure in the disk and corona and a highly relativistic,
    Poynting-dominated funnel jet in the polar regions.  The polar jet
    is nearly consistent with the stationary paraboloidal
    Blandford-Znajek model of an organized field threading the polar
    regions of a rotating black hole.  How can a disordered accretion
    disk and corona lead to an ordered jet?  We show that the polar
    jet is associated with a strikingly simple angular-integrated
    toroidal current distribution $dI_\phi/dr \propto r^{-5/4}$, where
    $I_\phi(r)$ is the toroidal current enclosed inside radius $r$. We
    demonstrate that the poloidal magnetic field in the simulated jet
    agrees well with the force-free field solution for a non-rotating
    thin disk with an $r^{-5/4}$ toroidal current, suggesting rotation
    leads to negligible self-collimation.  We find that the polar
    field is confined/collimated by the corona. We also study the
    properties of the bulk of the simulated disk, which contains a
    turbulent magnetic field locked to the disk's Keplerian rotation
    except for rapidly rotating prograde black holes ($a/M\gtrsim
    0.4$) for which within $r\lesssim 3GM/c^2$ the field locks to
    roughly half the black hole spin frequency. The electromagnetic
    field in the disk also scales as $r^{-5/4}$, which is consistent
    with some Newtonian accretion models that assume rough
    equipartition between magnetic and gas pressure.  However, the
    agreement is accidental since toward the black hole the magnetic
    pressure increases faster than the gas pressure. This field
    dominance near the black hole is associated with magnetic stresses
    that imply a large effective viscosity parameter $\alpha\sim 1$,
    whereas the typically assumed value of $\alpha\sim 0.1$ holds far
    from the black hole.

  \end{abstract}




  \section{Introduction}\label{introduction}

  Black hole accretion is one of the most powerful sources of energy
  in the universe.  A substantial fraction of the gravitational
  binding energy of the accreting gas is released within tens of
  gravitational radii from the black hole, and this energy supplies
  the power for a variety of astrophysical systems including active
  galactic nuclei, X-ray binaries, and gamma-ray bursts.  Elucidating
  the processes that take place in the central regions of black hole
  disks is obviously crucial if we wish to understand the physics of
  these energetic objects.

  Magnetized, differentially-rotating accretion disks exhibit the
  magneto-rotational instability (MRI) and magnetohydrodynamic
  turbulence \citep{bh91,bh98}, which generate large spatio-temporal
  variations in all fluid quantities and strong correlations between
  fluid quantities.  Recent general relativistic magnetohydrodynamic
  (GRMHD) simulations of black hole accretion systems have begun to
  resolve these processes and have revealed a flow structure that can
  be decomposed into a disk, corona, disk wind, and highly magnetized
  polar region that contains a jet \citep{dhk03,mg04}. As expected,
  the simulations show complex time-dependent behavior in the disk,
  corona, and wind. Surprisingly, however, the polar regions of the
  flow are found to have a simple structure with a nearly force-free,
  time-steady Poynting-dominated jet
  \citep{mg04,hk06,mckinney2006c}. The numerical solution here is
  quantitatively consistent with the relativistic force-free model
  proposed by \citet{bz77}, hereafter BZ \citep{mg04}.

  The primary question this paper explores is the following: how can a
  turbulent accretion disk lead to a nearly stationary, collimated and
  ordered Poynting-dominated jet?  In simple force-free models of the
  disk-jet coupling, like the one developed by BZ, one finds
  stationary solutions for a fixed toroidal current and angular
  velocity of the disk (see, e.g. \citealt{bz77}).  Since in GRMHD
  simulations the poloidal field threading the black hole is simple
  and nearly stationary, this implies that the toroidal current must
  also be simple and nearly stationary. However, it has not yet been
  known or understood what {\it radial dependence} would be chosen by
  turbulent accretion flows driven by the MRI.  In order to place the
  GRMHD simulation results in the context of analytical models of jets
  and winds that treat the disk as a equatorial boundary condition,
  our first objective is to determine the radial dependence of the
  toroidal current, angular velocity of the plasma, and angular
  velocity of the magnetic field.

  Given the radial distribution of the toroidal current and angular
  velocities of the plasma and field, one can generate force-free
  models of the jet that approximate the disk as an infinitely thin
  rotating conductor. General relativistic force-free solutions of
  this kind are obtained and then compared to GRMHD simulations and
  the models of BZ in a followup paper \citep{mn06}.

  Our second objective in this paper is to determine the angular
  distribution of toroidal currents and the flow pattern of poloidal
  currents in the accretion flow. The location of the toroidal
  currents helps identify the role played by the weakly magnetized
  corona in confining the highly magnetized jet. For example, the
  corona provides forces that balance the forces due to the strong
  poloidal field gradients (associated with strong toroidal currents)
  at the boundary between the magnetized jet and corona.  Hence, the
  corona can be understood as required to confine the magnetized jet.
  Also, a stationary model must have {\it poloidal} currents that
  close like a circuit, yet it has not been known where such poloidal
  currents flow in MHD turbulent disks around black holes.  We study
  the distribution of poloidal currents that are associated with the
  outgoing power of the jet/wind in order to establish where the
  poloidal currents flow and to resolve the issue of current closure.

  Our third objective is to check if either of the two magnetic field
  geometries described by BZ, viz., the split-monopole and the
  paraboloidal geometries (originally discovered by \citealt{michel73}
  and \citealt{blandford76}, respectively), is a good description of the
  jets found in our GRMHD simulations. We find that neither model is
  satisfactory.  Instead we identify a third model, in between the other
  two and close to the paraboloidal model, which agrees surprisingly
  well with the simulations as long as the jet is nearly force-free.

  The toroidal current in the split-monopole and paraboloidal
  solutions scale with radius as $dI_\phi/dr \propto r^{-2}, ~r^{-1}$,
  respectively, whereas the toroidal current in the GRMHD simulations
  is found to scale as $r^{-5/4}$.  Interestingly, the latter scaling
  is identical to that proposed by \citet{bp82} (BP), who developed a
  non-relativistic self-similar magnetohydrodynamic (MHD) model of
  disk winds by assuming that the sound speed and \alf~speed in the
  disk scale similarly with radius. Our fourth objective in this paper
  is to establish how the disk magnetic field strengths, sound speed,
  \alf~speed, plasma speed depend on radius in order to determine
  whether the agreement between the GRMHD simulations and the BP model
  has a deep physical significance or is merely a coincidence.  The
  answer appears to be the latter in the sense that the assumptions
  made by BP are broken near the black hole.  More importantly, the
  field threading the disk is disorganized and the disk wind is
  thermally-driven instead of behaving like a ``bead on a wire'' as in
  the BP model.

  Prior studies of the BZ power output suggested that the black hole
  power output should be too small compared to the disk power output
  and too small to account for the most powerful radio sources
  \citep{ga97,lop99}.  Such studies assumed that the effective
  viscosity parameter $\alpha$ was as determined in non-relativistic
  simulations and assumed the field strength near the black hole was
  set by sub-equipartition arguments.  Our final objective is to
  determine the magnetic $\alpha$ viscosity parameter as a function of
  radius within the disk.  Both of their assumptions end up not
  applying near the black hole.

  \section*{Paper Outline}

  In section~\ref{GRMHDDISK}, we discuss the origin of the ordered
  poloidal field in GRMHD simulations.  We show that the
  angular-integrated toroidal currents in the turbulent accretion disk
  follow a simple power-law behavior.  We discuss the angular
  structure of the toroidal currents and the flow of poloidal currents
  in the accretion flow.  We discuss the field angular velocity in the
  transition region between the accretion disk and black hole. In
  section~\ref{emproperties}, we study the GRMHD accretion flow in
  order to extract other electromagnetic properties, such as the
  magnetic field strength as a function of radius.  We test the
  assumptions of BP against our GRMHD numerical models and study the
  electromagnetic stress that leads to an enhanced angular momentum
  transport near the black hole.  In section~\ref{limitations}, we
  discuss the limitations of our calculations.  Finally, in
  section~\ref{conclusions}, we discuss our results and conclude.

  In appendix \ref{GRMHD}, we summarize the GRMHD equations of motion
  and point out the reduction to the force-free set of equations.  In
  appendix \ref{reducegrmhd}, we show how to obtain force-free solutions
  in Schwarzschild and flat spacetimes for an arbitrary current sheet at
  the equator, and we discuss how the disk currents are integrated to
  obtain a toroidal current density as a function of radius.  We also
  give three example solutions corresponding to the split-monopole,
  paraboloidal, and our new self-similar solution.

  \section*{Units and Notation}

  The units in this paper have $G M = c = 1$, which sets the scale of
  length ($r_g\equiv GM/c^2$) and time ($t_g\equiv GM/c^3$).  The
  horizon is located at $r=r_+\equiv r_g(1+\sqrt{1-(a/M)^2})$).  For a
  black hole with angular momentum $J=a GM/c$, $a/M$ is the
  dimensionless Kerr parameter with $-1\le a/M \le 1$.  In order to
  obtain a density for a given mass accretion rate, one requires the
  field as a function of black hole spin given by GRMHD models such as
  described in
  \citet{mckinney2005a,mckinney2005b,mckinney2005c,mckinney2006c}.
  The mass scale is determined by setting the observed
  (model-dependent measured or inferred) mass accretion rate
  ($\dot{M}_0$) equal to the accretion rate through the black hole
  horizon as measured in a simulation.  So the mass scale is set by
  the mass accretion rate ($\dot{M}_0$) at the horizon, such that
  $\rho_{0,disk}\equiv \dot{M}_0[r=r_+] t_g/r_g^3$ and the mass scale
  is then just $m\equiv \rho_{0,disk} r_g^3 = \dot{M}_0[r=r_+] t_g$.

  The results of the simulations can be applied to any astrophysical
  system once the value of $\rho_{0,disk}$ is estimated.  For example,
  a collapsar model with $\dot{M}=0.1\msun s^{-1}$ and $M\approx
  3\msun$ has $\rho_{0,disk}\approx 3.4\times 10^{10}{\rm g}\cm^{-3}$
  \citep{mw99}.  M87 has a mass accretion rate of $\dot{M}\sim
  10^{-2}\msun\yr^{-1}$ and a black hole mass of $M\approx 3\times
  10^9\msun$ \citep{ho99,reynolds96} giving $\rho_{0,disk}\sim
  10^{-16} {\rm g}\cm^{-3}$.  GRS 1915+105 has a mass accretion rate
  of $\dot{M}\sim 7\times 10^{-7}\msun\yr^{-1}$ \citep{mr94,mr99,fb04}
  with a mass of $M\sim 14\msun$ \citep{greiner2001a} (but see
  \citealt{kaiser04}).  This gives $\rho_{0,disk}\sim 3\times
  10^{-4}{\rm g}\cm^{-3}$.

  The notation follows \citet{mtw73} and the signature of the metric
  is $-+++$.  Tensor components are given in a coordinate basis.  The
  components of the tensors of interest are given by $g_{\mu\nu}$ for
  the metric, $F^{\mu\nu}$ for the Faraday tensor, $\dF^{\mu\nu}$ for
  the dual of the Faraday, and $T^{\mu\nu}$ for the stress-energy
  tensor. The determinant of the metric is given by $\detg \equiv {\rm
  Det}(g_{\mu\nu})$. The field angular frequency is $\Omega_F\equiv
  F_{tr}/F_{r\phi}=F_{t\theta}/F_{\theta\phi}$.  The magnetic field
  can be written as $B^i=\dF^{it}$.  The poloidal magnetospheric
  structure is defined by the $\phi$-component of the vector potential
  $A_\phi$. A stationary, axisymmetric current system is defined by
  the current density $\mathbf{J}$ and the enclosed (from the pole to
  some point) poloidal current ($B_\phi\equiv\dF_{\phi t}$).  The
  electromagnetic luminosity is $L\equiv -2\pi \int_\theta d\theta
  {T^{(EM)}}^r_t r^2\sin\theta$. See
  \citet{gmt03,mg04,mckinney2004,mckinney2005b,mckinney2005c,mckinney2006a}
  for details on this standard notation.

  \section{The Organized Polar Field}\label{GRMHDDISK}

  In this section we discuss the origin and nature of the organized
  field threading the black hole.  We first review some relevant
  results from GRMHD simulations of accretion flows.  We then
  demonstrate that at large radii the jet from the black hole is
  electromagnetically pure, while the disk wind is dirty.  Next, we
  show that the angular-integrated toroidal current over the accretion
  flow is simple and has a power-law radial dependence, which is
  associated with the simple organized polar field.  We analyze the
  angular structure of these toroidal currents to locate the toroidal
  currents (or equally the large poloidal field gradients). We show
  how the jet and disk wind are associated with poloidal currents that
  close in an electric circuit.  Next, we determine the radial
  dependence of the disk-averaged plasma and field angular velocities,
  which have a simple behavior.  This also elucidates how strongly the
  field in the disk couples to the rotation of the black hole.
  Finally, we compare the polar field in the GRMHD simulations to the
  non-rotating force-free field solution that emerges from the same
  power-law toroidal current as in the GRMHD simulations.  The
  agreement found between these models demonstrates that the polar jet
  is accurately described by a force-free model and that rotation
  plays a negligible role in self-collimating the polar jet.  Rotation
  may still play a role in {\it indirectly} collimating the jet via
  the rotation of the surrounding coronal wind.

  \subsection{Review of Prior GRMHD Simulation Results}

\begin{figure}
    \includegraphics[width=3.3in,clip]{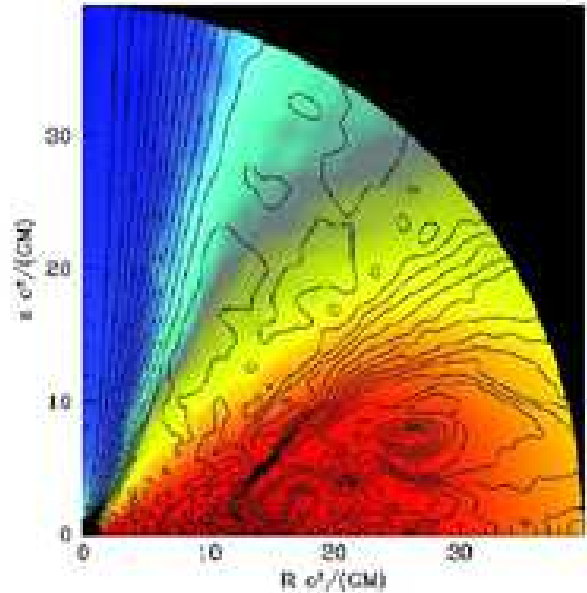}

    \caption{Time-averaged poloidal magnetic field (solid black lines)
      for the fiducial GRMHD numerical model with $a/M=0.9375$ studied
      by \citet{mg04}.  Color shows the logarithm of the time-averaged
      rest-mass density ($\log \rho_0$) from highest (red) to lowest
      (blue) densities.  The black hole is located at $(R,z) = (0,0)$.
      Notice (i) the ordered, magnetized, low-density jet near the
      polar axis, (ii) the turbulent, disordered disk in the
      equatorial region, and (iii) the coronal region in between with
      only a weak disordered field. }
    \label{grmhdfield} \end{figure}

  In this section, we review some relevant results from GRMHD
  simulations, where we start the discussion with the fiducial model
  of \citet{mg04}.  Figure~\ref{grmhdfield} shows the time-averaged
  field geometry for this fiducial model.  The black hole has a spin
  of $a/M=0.9375$, which is close to the equilibrium value of $a/M\sim
  0.92$ \citep{gsm04}.

  The simulated accretion flow is dominated by a turbulent
  hydromagnetic dynamo driven by the MRI, which drives an increase in
  poloidal field strength whose saturated magnitude is insensitive to
  the initial poloidal field strength \citep{dhk03,mg04}.  The dynamo
  eventually dies out unless modelled in three dimensions
  \citep{cowling34}. For the two-dimensional simulations of
  \citet{mg04}, measurements are made only during the turbulent period
  and their results are consistent with three-dimensional simulations.

  The polar region contains a well-ordered field whereas the disk
  appears to have a turbulent and disordered field
  \citep{mg04,hirose04,mckinney2005a}. Between these two regions is
  the corona which, in a time-averaged sense, contains only weak
  disordered fields.  In the figure, the field that appears to come
  from the disk does not reach large distances, whereas the organized
  field in the polar region reaches large radii \citep{mckinney2006c}.

  It is important to note that the organized polar field geometry
  shown in Figure~\ref{grmhdfield} is generic for simulations that are
  initialized with a well-organized poloidal field with or without net
  flux. Here an ``organized'' poloidal field means a field geometry
  with few poloidal polarity changes, but it could be initially
  contained entirely within the disk. The quasi-stationary structure
  of the accretion flow and flux threading the black hole or disk
  otherwise depends little on the initial field strength or geometry
  \footnote{ The codes used in the studies mentioned above preserve
  the solenoidal constraint ($\nabla\cdot\mathbf{B}=0$) to machine
  accuracy, and so the codes preserve the net poloidal flux in the
  system.  Hence, the net radial flux remains constant and the
  $\theta$ flux only changes by losing flux into the black hole or
  through the outer radial boundary.  Codes that violate this property
  might generate large artificial net flux and then regions of
  organized flux could not be trusted.} \citep{mg04,hirose04}.

  Cases when the black hole does not end up with an organized field
  include an initially purely toroidal field (no toroidal currents so
  no poloidal field) \citep{dv05} or highly tangled field (McKinney \&
  Gammie, in prep.).  With a purely toroidal field, no jet is
  produced.  With a highly tangled field, the Poynting-dominated jet
  does not form since it is continuously contaminated with disk
  material, and so the corona and coronal wind dominate the entire
  polar region. Even when the disk is initially threaded by net flux,
  the quasi-stationary corona still only contains weak disordered
  field.

  The lack of organized field in the corona or threading the disk
  might be understandable as a result of the inability of flux to be
  simply advected radially inward if the ratio of viscosity to
  magnetic diffusivity (magnetic Prandtl number $P_m$) is such that
  $P_m\lesssim H/R$, where $H/R$ is the pressure scale-height per unit
  radius of the disk (see discussion in, e.g., \citealt{lop99},
  sections 3.1 and 3.2).  In thin turbulent-driven accretion flows,
  the magnetic Prandtl number is order unity, and so one does not
  expect to be able to simply accrete net flux from large radii.  On
  the other hand, thick advection-dominated accretion flows are
  expected to be able to advect a nonnegligible flux.  In
  three-dimensional models, the advection of flux may occur in
  isolated flux tubes and still build near the black hole
  \citep{su05}.  The physics of how large-scale flux can be accreted
  through disks remains an open issue (see, e.g.,
  \citealt{rgb06,cdc06}).

  So in the GRMHD simulations, how does the black hole become threaded
  with flux and how is the magnitude of the field maintained?  Does
  large-scale flux simply advect inward despite the above-mentioned
  issues?  In the simulations, the initial large-scale flux around the
  black hole is created as a result of the accretion of equatorial
  field loops that do not thread vertically through the disk, thus
  bypassing the problem of advecting large-scale flux threading the
  entire disk.  The long-term flux is maintained by the 2D
  axisymmetric dynamo that drives single polarity poloidal loops
  (corresponding to single polarity toroidal currents) to twist in
  an axisymmetric sense, break through reconnection, and interchange
  around each other within the disk allowing the black hole to accrete an
  arbitrary poloidal polarity\footnote{In a 3D dynamo non-axisymmetric
  modes would more vigorously generate such varying polarities and
  allow direct radial interchange not allowed in axisymmetry.}.  Near the
  black hole the electromagnetic field dominates and this allows the
  magnetic flux threading the horizon to grow by the attraction of
  similarly-signed toroidal currents and the repulsion of
  oppositely-signed toroidal currents -- a classical electrodynamical
  phenomena.

  The magnetic field strength near the black hole saturates when
  material forces of the disk+corona can just support the magnetic
  pressure of the polar field. Thus one expects the magnitude of the
  magnetic pressure to be in near equipartition with the magnitude of
  the gas pressure fairly close to the horizon, which is the case
  \citep{mg04}. This balance is qualitatively similar to the balance
  that is associated with the ``magnetospheric radius'' in accreting
  neutron star systems.

  Also important is the accretion of large-scale polarity changes that
  can overwhelm any prior field built-up around the black hole.  The
  relevance of accreting large-scale opposite polarities probably
  depends on the astrophysical system (see, e.g.,
  \citealt{narayan2003}).  Reconnection tends to convert the
  pre-existing flux into thermal energy and erase the organization of
  the polar field. When the reconnection time-scale is long, the field
  energy associated with the flux is difficult to remove once the flux
  is in place.

  The magnetized jet is produced as the black hole spin and disk rotation create
  a large toroidal field whose gradient launches a significant
  fraction of accreted flux out along the poles.  This process
  effectively leads to net flux threading the black hole since the
  flux that reaches large distances becomes causally disconnected from
  the disk \citep{mckinney2006c}.  The stronger the black hole spin,
  the stronger is this effect.  In a stationary state, the organized
  polar field threads the black hole horizon, but {\it not} the inner
  disk \citep{hirose04,mckinney2005a}.

  In summary, the black hole naturally becomes threaded by organized
  field when the disk contains a field with few large-scale poloidal
  polarity changes.  Early in the simulation, the field threading the
  black hole is advected through the equatorial region rather than
  being advected as large-scale flux threading the entire disk.  The
  magnetic field strength near the black hole saturates when a balance
  is reached between the polar magnetic pressure and the disk+corona
  pressure.  The disk and black hole rotations drive the flux near the
  black hole to large radii, and this effectively leads to a net flux
  threading the horizon.  In a quasi-stationary state, large-scale
  field threads the hole but not the inner disk.

  \subsection{Power Output of Black Hole and Disk}\label{powercompare}

  In principle, a significant fraction of the electromagnetic power
  may come from a disk wind \citep{bz77} and may even dominate the
  black hole power output \citep{ga97,lop99}.  However, the lack of an
  organized field threading the disk suggests that the electromagnetic
  output of the disk may be seriously compromised compared to those
  simple models that have treated the disk as just a boundary
  condition. To test these ideas, we compare the electromagnetic power
  output of the black hole and the disk in the fiducial ($a/M=0.9375$)
  model shown in Figure~\ref{grmhdfield}.

  \begin{figure}
    \includegraphics[width=3.3in,clip]{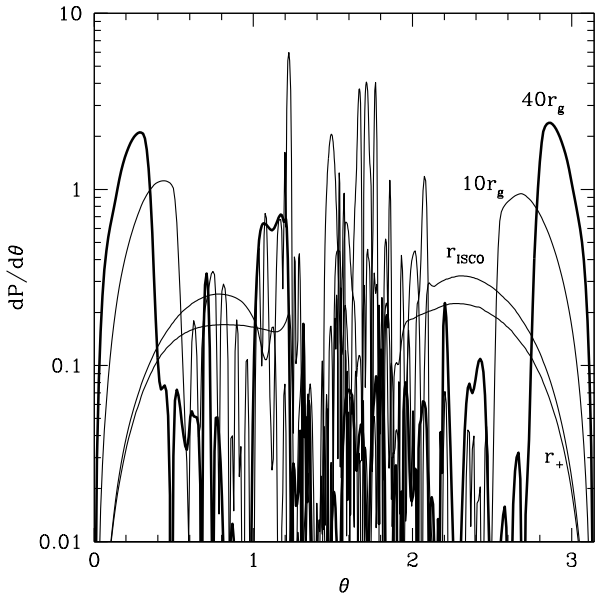}

    \caption{The angular density of the electromagnetic power $\langle
      dP/d\theta\rangle$ vs. $\theta$ in radians at four selected radii,
      $r=\{r_+,~r_{\rm ISCO},~10r_g,~40r_g\}$, for the fiducial GRMHD
      model with $a/M = 0.9375$ described in \citet{mg04}. The thick solid
      line corresponds to $r=40r_g$, the outer radius of the computational
      domain for this simulation. Only the black hole electromagnetic
      power output survives at this distance, whereas the disk
      electromagnetic power output has been efficiently converted into
      other forms. } \label{energyfluxgrmhd}
  \end{figure}

  Let us define the time-averaged angular density of the
  electromagnetic power output as
  \begin{equation}
    \left\langle\frac{dP}{d\theta}\right\rangle = 2\pi r^2
    \left\langle-{T^{(EM)}}^r_t\right\rangle , \label{dPdtheta}
  \end{equation}
  where $-{T^{(EM)}}^r_t$ is the radial electromagnetic flux written
  in a suitable coordinate basis (either Boyer-Lindquist or
  Kerr-Schild coordinates). The time-averaging is performed over
  approximately $8$ orbital periods at $r=10r_g$ once the flow reaches
  a quasi-stationary, turbulent state. Figure~\ref{energyfluxgrmhd}
  shows $\langle dP/d\theta\rangle$ as a function of $\theta$ at four
  radii: the horizon at $r=r_+$, the inner-most stable circular orbit
  (ISCO) at $r=r_{\rm ISCO}$, $r=10r_g$, and $r=40r_g$. Notice how
  smooth and well-behaved the power is near the poles and how erratic
  and disordered it is away from the poles.












\begin{table*}
\begin{center}
\begin{tabular}{lllllll}
\hline
\multicolumn{7}{|c|}{\bfseries GRMHD Model Electromagnetic Power} \\
\hline
$a$ & $P[r_+]$ & $P[r_{\rm ISCO}]$ & $P[r=10r_g]$  & $P[r=40r_g]$ & $\Gamma[r=5\times 10^3r_g]$ & $\theta_j[r=5\times 10^3r_g]$ \\

0.9375 & 0.324  &    0.407 &     0.595   &   0.374 & 10 & $5^\circ$ \\






\hline
\end{tabular}
\caption{Electromagnetic power (per unit $(B^r)^2$ on horizon at
poles), Lorentz factor, and half-opening angle. }
\label{tbl1}
\end{center}
\end{table*}

  The total electromagnetic power at any radius $r$ is given by
  \begin{equation}
    P(r) = \int_0^\pi \left\langle\frac{dP}{d\theta}\right\rangle
    \sin(\theta) d\theta .
  \end{equation}
  Table~\ref{tbl1} lists values of $P$ at the same four radii shown in
  Figure~\ref{energyfluxgrmhd}.  For comparison with the force-free
  models discussed in \citet{mn06}, the power is normalized such that
  \begin{equation}
    P(r) \rightarrow \frac{P(r)}{((B^r)^2|_{\theta=0}) 4\pi r_g^2 c} ,
  \end{equation}
  where the field is in Gaussian units and is measured on the horizon
  at the poles at the final time of the simulation.

  At the horizon the electromagnetic power is that from the black hole
  with most of the net electromagnetic power coming from the polar
  regions.  With increasing radius, electromagnetic power from the
  disk is added.  However, this power is steadily converted into
  matter energy so that, by $r=40r_g$, most of it is no longer in
  electromagnetic form.  This explains why the power at $r=40r_g$ in
  Table~\ref{tbl1} is not much larger than the power at $r=r_+$.

  Table~\ref{tbl1} also gives the Lorentz factor $\Gamma$ of the jet
  far from the black hole and the half-opening angle $\theta_j$ of the
  jet (defined as the angle at which $\langle dP/d\theta\rangle$ is
  maximum).  These results are from \citet{mckinney2006c}.

  In summary, we find that the electromagnetic power at the poles of
  the black hole remains undiminished out to the largest radius shown,
  whereas the electromagnetic power from the disk region decreases
  outward as it is efficiently converted into kinetic and thermal
  energy of the plasma.  Simple estimates of the power output of the
  black hole and disk did not consider the conversion of
  electromagnetic power into thermal and material power in the corona
  and disk wind \citep{ga97,lop99}, and so they may have seriously
  overestimated the power from the disk.  Thus, the black hole may
  dominate the electromagnetic power output of accretion systems.
  Also, when the disk is turbulent, the black hole remains the only
  possible clean ($b^2/(\rho_0 c^2)\gg 1$) source of electromagnetic
  power.

  \subsection{A Power-Law Radial Distribution of Toroidal Current}

  For stationary flows, the toroidal current directly leads to the
  poloidal field structure.  Since the Poynting-dominated jet has a
  simple, stationary poloidal structure, there must be simple toroidal
  currents to support the field.  However, given the turbulence in the
  disk and the efficient conversion of electromagnetic power from the
  disk into material power, one might assume that the electromagnetic
  properties of the disk would be complex and hard to relate in any
  simple way to the organized polar flux. This is certainly suggested
  by Figures~\ref{grmhdfield} and~\ref{energyfluxgrmhd}.

  As discussed in the introduction, the motivation for studying the
  toroidal current comes from simplified force-free models that
  include the accretion disk as an equatorial boundary condition, such
  as the paraboloidal BZ model \citep{bz77}.  In general, models of
  winds and jets typically specify the toroidal current in the disk
  (or stellar surface) and find the corresponding solution (see, e.g.
  \citealt{michel73,oka74,oka78,blandford76,bp82,lov86,hey89,nit91,li92,appl92,appl93,bes93,conone94,con94,con95a,con95b}).
  We stress that for a stationary solution the toroidal current and
  poloidal field are just different languages for the same physics,
  but our motivation is to see if a turbulent GRMHD disk can be
  related to the vast array of models that describe the disk as simply
  a current sheet.

  The relevant questions are: 1) Are such simple models applicable to
  thick, turbulent, magnetized accretion flows?  ; 2) What {\it radial
  dependence} of the toroidal current is the ``correct'' choice?  ;
  and 3) Does the angular-integrated toroidal current predict the
  shape of the organized polar field?

  Let us consider the angular-integrated (over {\it all} angles)
  toroidal current that for a stationary flow corresponds to a line
  integral of the magnetic field around a closed poloidal loop.  We
  integrate over all angles to capture currents that sometimes rise
  into the corona and to capture variations in the disk+corona
  thickness in time and as a function of radius.  We consider the
  angular distribution of toroidal currents in the next section.

  The current density (as given by equation (\ref{currentdensity})) can
  be integrated to determine the net toroidal current enclosed within
  the volume between $r_0$ and $r$.  This invariant current is
  \begin{eqnarray}\label{iphiintegral}
    I_{\phi} & = & \int_{r_0}^{r}\int_{\theta=0}^\pi \left(J^\mu
    d\Sigma_\mu\right)\nonumber\\
    & = &\int_{r_0}^{r}\int_{\theta=0}^\pi \detg \left(J^\phi dr' d\theta' \right)\nonumber\\
    & \equiv & \int_{r_0}^{r} \left(\frac{dI_{\phi}}{dr'} dr' \right) ,
  \end{eqnarray}
  where $d\Sigma_\mu \equiv \epsilon_{\mu\nu\alpha\beta} t^\nu
  dr^\alpha d\theta^\beta$, $t^\mu=\{1,0,0,0\}$ is the time-like
  Killing vector, and $dI/dr$ is the toroidal current per unit radius.
  Note that the magnitude of the enclosed current is set by the value
  of $I_\phi(r_0)$, which is an arbitrary constant and set to be the
  enclosed current in the numerical grid of size $dr$ at $r_0$. Only
  the magnitude of the current density has physical significance.
  Notice that
  \begin{equation}\label{didr}
    \frac{dI_{\phi}}{dr} \equiv \int_{\theta=0}^\pi \detg
    J^\phi d\theta ,
  \end{equation}
  where $\detg\approx r^2\sin\theta$ far from the black hole.

  In the following we are particularly interested in models with a
  power-law dependence of the current density, i.e.,
  \begin{equation}\label{powerlawcurrent}
    \frac{dI_{\phi}}{dr} \propto \frac{1}{r^{2-\nu}} .
  \end{equation}
  We are motivated by the fact that the split-monopole and
  paraboloidal force-free models both have currents of this form, with
  $\nu=0$ and 1, respectively.  Example solutions are given in
  appendix~\ref{reducegrmhd}.

  \begin{figure}

    \begin{center}
      \includegraphics[width=3.3in,clip]{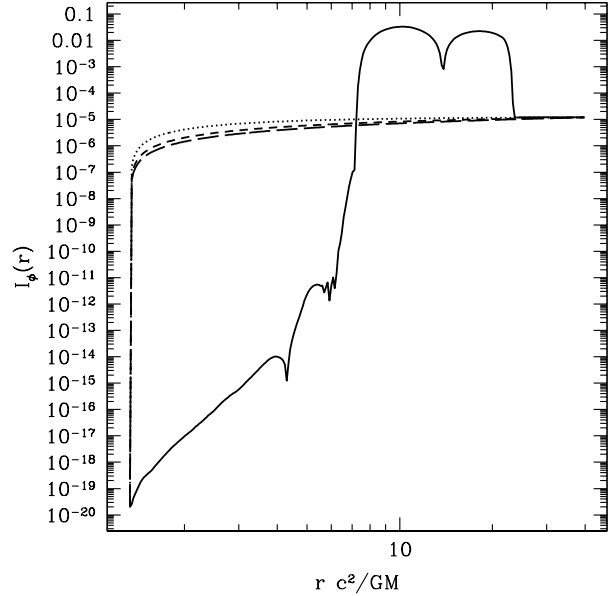}
    \end{center}

    \caption{Vertically-integrated enclosed toroidal current as a
      function of radius.  Solid line shows the current in the GRMHD
    numerical model with $a/M=0.9375$ at $t=50t_g$ (nearly the initial
    conditions).  The dotted line is the $\nu=0$ monopole solution,
    the long-dashed line is the $\nu=1$ paraboloidal solution, and the
    short-dashed line is the $\nu=3/4$ solution. Clearly the initial
    conditions do not follow any simple power-law behavior. }
    \label{torencplotgrmhdt0}
  \end{figure}

  \begin{figure}

    \begin{center}
      \includegraphics[width=3.3in,clip]{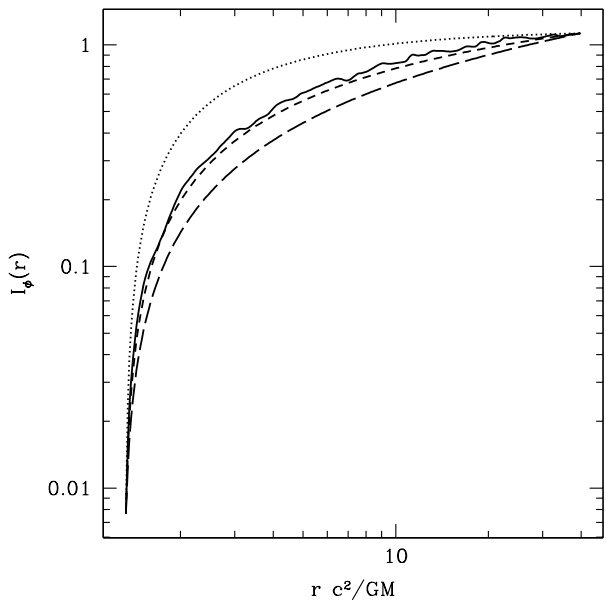}
    \end{center}

    \caption{Similar to Figure~\ref{torencplotgrmhdt0}, except that
      the solid line shows the result for the GRMHD numerical model at
      $t=1000t_g$, during a period of strong sustained disk turbulence.
      By the action of the magneto-rotational instability, the GRMHD
      model's toroidal current has redistributed itself to
      closely follow a $\nu=3/4$ power-law dependence. }
    \label{iencvsrgrmhd9375_d20}
  \end{figure}

  \begin{figure}
    \begin{center}
      \includegraphics[width=3.3in,clip]{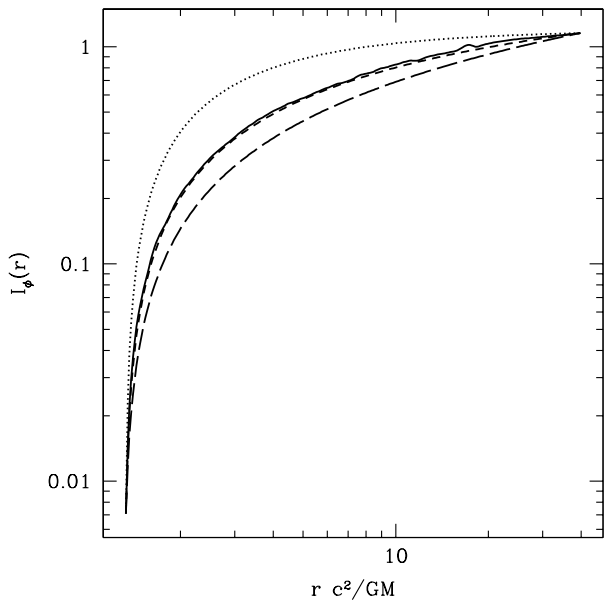}
    \end{center}

    \caption{Similar to Figure~\ref{iencvsrgrmhd9375_d20}, except the
      enclosed toroidal current has been time-averaged over several
      turbulent dynamical times at $r\sim 10r_g$.  Evidently, the
      simple $\nu=3/4$ power-law behavior for the toroidal current
      holds for both the time-averaged quasi-stationary turbulent
      state {\it and} at every moment in time once the turbulence has
      grown to saturation. }
    \label{iencvsrgrmhd9375}
  \end{figure}

  The solid line in Figure~\ref{torencplotgrmhdt0} shows the enclosed
  toroidal current $I_\phi(r)$ at a very early time ($t=50t_g$) of the
  fiducial simulation described in \citet{mg04}.  At this time, the
  system has hardly deviated from the initial conditions and there is
  very little accretion taking place.  Correspondingly, the current
  has a fairly complicated dependence, which primarily reflects the
  particular initial conditions chosen for this simulation.

  Figure~\ref{iencvsrgrmhd9375_d20} shows the enclosed toroidal
  current at a later time ($t=1000t_g$) when the accretion flow is
  highly turbulent and has reached a quasi-steady state.  We see that
  the current distribution has changed dramatically from its initial
  distribution.  More importantly, the current profile looks smooth
  and simple.  Figure~\ref{iencvsrgrmhd9375} shows the enclosed
  current as time-averaged over the period $t=500t_g$ to $1500t_g$
  (roughly a turbulent time scale at $r=40r_g$).  The result is very
  similar to that shown in Figure~\ref{iencvsrgrmhd9375_d20}, except
  that the current looks even smoother because of the averaging.

  In Figures~\ref{torencplotgrmhdt0}-\ref{iencvsrgrmhd9375}, we show for
  comparison the enclosed currents corresponding to the split-monopole
  ($\nu=0$, dotted line) and paraboloidal ($\nu=1$, long-dashed line)
  models.  It is clear that neither of these power-law models is a good
  representation of the enclosed current in the steady state GRMHD
  model.  On the other hand, the short-dashed lines, which correspond to
  a power-law model with $\nu=3/4$, describe the quasi-stationary GRMHD
  results surprisingly well.  This particular model is associated with a
  radial dependence of the current density of the form
  $dI_{\phi}/dr\propto r^{-5/4}$.

  A few interesting conclusions can be reached from these results: 1)
  The GRMHD model is nearly coincident with the $\nu=3/4$ model; 2)
  The GRMHD model is not consistent with the split-monopole or
  paraboloidal models; 3) Despite the complicated nonlinear
  turbulence, the currents in the disk are simple not only on average,
  but at {\it each} moment in time.

  The simple $\nu=3/4$ current distribution described above is
  entirely consistent with the smooth field distribution seen in the
  evacuated polar region in Figure~\ref{grmhdfield}.  It is also
  consistent with the fact that the Poynting-dominated jet is nearly
  stationary and nearly resembles BZ's paraboloidal solution
  \citep{mg04,mckinney2006c}.  Furthermore, the fact that the best fit
  is obtained for $\nu=3/4$ rather than $\nu=1$ explains why the
  large-scale field lines in the jet are nearly paraboloidal but
  somewhat less collimated in the GRMHD simulations
  \citep{mckinney2005c,mckinney2006c}.  Section~\ref{overlay}
  discusses this point further.

  The above results are roughly independent of the initial field
  geometry assumed in the GRMHD simulations.  \citet{mg04} considered
  different initial conditions such as multiple magnetic loops in the
  initial torus, a net vertical field, and loops of alternating
  poloidal direction.  Once these simulations have reached a
  quasi-steady state, each closely follows a power-law toroidal
  current density with $\nu=3/4$.  This is despite the fact that the
  particular angular structure of the toroidal current varies
  considerably.  For example, for thin disks with alternating polarity
  of multiple field loops, there is no strong poloidal field in the
  funnel, but the toroidal current still maintains the $\nu=3/4$
  power-law dependence.

  We also investigated the dependence of the toroidal current
  distribution on the black hole spin.  Once again we find that, for
  all $a/M$ ranging from $-0.999$ to $+0.999$, the toroidal current
  settles down to a $\nu=3/4$ power-law distribution once the flow
  reaches quasi-steady state.  This is despite significant changes in
  the enclosed poloidal current as shown in
  Figure~\ref{bphiaphigrmhd}, since for slowly spinning black holes
  there is negligible poloidal current flowing in the funnel region.

  How robust are these results to changes in the mass distribution and
  disk thickness?  The fiducial model studied has an initial mass
  distribution of a hydrostatic equilibrium torus with a constant
  specific angular momentum such that $H/R\sim 0.3$.  An alternative
  initial mass distribution we tried has a quasi-equilibrium Keplerian
  disk with a Gaussian vertical mass distribution with $H/R\sim 0.3$.
  Once the turbulence reaches the nonlinear phase, this alternative
  model also has a toroidal current closely matching the $\nu=3/4$
  profile.

  We have also studied a thin (i.e. small $H/R$) Keplerian disk with an
  initial Gaussian vertical distribution with an ad hoc cooling model to
  keep $H/R\sim 0.05$, and we find that this model also obeys the
  $\nu=3/4$ toroidal current distribution.  The only qualitative change
  is that there is a larger variance around the $\nu=3/4$ solution.
  However, the solution is still quite different from the monopole or
  paraboloidal solutions.

  \subsection{Angular Structure of Toroidal Current}

  Of course, the angular-integrated toroidal current does not reveal
  the angular location of the currents.  By integrating equation
  (\ref{iphiintegral}) over a finite radial range for each $\theta$,
  one can identify structures in the accretion flow with the toroidal
  currents.  While the instantaneous value of $I_\phi(\theta)$ is very
  oscillatory and difficult to interpret, the running integral
  ($\int_0^\theta I_\phi(\theta')d\theta'$) is relatively smooth so we
  focus on this quantity.

\begin{figure}
    \begin{center}
      \includegraphics[width=3.3in,clip]{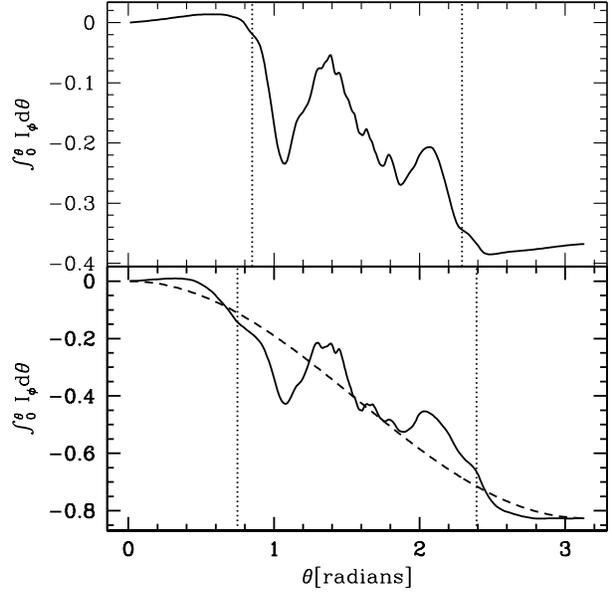}
    \end{center}

    \caption{Both panels show the running integral of the toroidal
      current $I_\phi$.  Top panel corresponds to the integral over
      the radial range from $r=r_+$ to $r=3r_g$.  Bottom panel
      corresponds to the integral over the radial range from $r_+$ to
      $r=10r_g$, where a rough fit is shown by the dashed line.  The
      vertical dotted lines show the location where $b^2/(\rho_0
      c^2)=1$ (i.e. the funnel-coronal interface). The overall
      toroidal current is seen to be distributed over a scale-height
      of the disk, but the funnel-corona and corona-disk interfaces
      harbor significant toroidal currents.  The polar field is
      associated with the toroidal current within the corona near the
      funnel. }
    \label{iphivstheta}
  \end{figure}

  We consider two radial sections of the accretion flow.  One radial
  range considered is $r=r_+$ to $r=3r_g$, while the other radial
  range considered is $r=r_+$ to $r=10r_g$.  Figure~\ref{iphivstheta}
  shows the radially integrated, angular running integral of
  $I_\phi(\theta)$ for the fiducial simulation in \citet{mg04}.  For
  this GRMHD simulation, the funnel jet extends from the poles inward
  by about $1/2$ to $1$ radian depending upon the radius.  Just beyond
  this range at the corona-funnel interface, the toroidal current
  changes sign and nearly annihilates itself in an integral sense.
  This toroidal current is associated with the polar field.  In the
  corona there is no organized field and the magnetic field is in
  equipartition with the gas pressure \citep{mg04}.  We find that
  across this coronal-funnel interface there is force balance between
  the organized field in the polar region and the equipartition,
  disorganized, weakly magnetized, hot plasma in the coronal region.
  If self-collimation is weak within the polar jet, then the corona is
  responsible for confining/collimating the polar jet.  The importance
  of rotation of the polar field is tested in later sections and in
  \citet{mn06}.

  The next jump in the toroidal current is at the corona-disk
  interface.  Within the disk the toroidal current oscillates with a
  slightly non-zero toroidal current.  For the integrated radial range
  of $r=r_+$ to $r=10r_g$, the angular integrated distribution of the
  toroidal current is roughly given by
  \begin{equation}
    \int_{\theta'=0}^\theta I(\theta') d\theta' \propto (1-\cos(\theta)) ,
  \end{equation}
  although there is significant sub-structure.  This fit is shown in
  Figure~\ref{iphivstheta}.

  While changes in the initial field geometry do not affect the radial
  distribution of toroidal currents with $dI_\phi/dr \propto
  r^{-5/4}$, the angular location of the toroidal currents (and so the
  location of poloidal field gradients) depends sensitively on the
  initial field geometry. Simulations that start with a highly
  disorganized field with many loop of different polarity lead to no
  organized flux in the poles and then the toroidal currents are
  primarily located inside the disk-corona interface.

  \subsection{Poloidal Currents and Current Closure}

  For stationary flows, the toroidal field is equally described by the
  poloidal current, and these are related to the power output of the
  jet.  At large radii, the radial electromagnetic power output is
  given by $\dot{E}\propto \Omega_F B_\phi B^r$, where $\Omega_F$ is
  the field rotation frequency, $B^r$ is the radial field strength,
  and $B_\phi$ is the ``polar enclosed poloidal current''
  ($B_\phi\approx R B^{\hat{\phi}}$).  In a stationary, axisymmetric
  flow, $\Omega_F$ and $B_\phi$ are constant along field lines
  labelled by the vector potential $A_\phi$ (also referred to as the
  flux function $\Psi$ or stream function).  Thus, the poloidal
  current is an indicator of the electromagnetic power output per unit
  poloidal field strength, and the closure of this poloidal current is
  best considered in a plot of $B_\phi$ vs. $A_\phi$.

\begin{figure}
    \begin{center}
      \includegraphics[width=3.3in,clip]{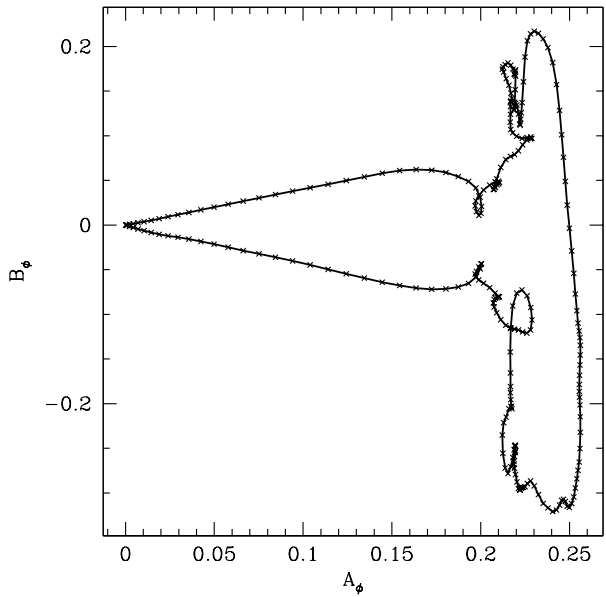}
    \end{center}

    \caption{Poloidal current enclosed from the pole ($B_\phi$) to the
      point given by the vector potential ($A_\phi$) for the surface
      given by spherical polar radius of $r=10r_g$. The value of
      $A_\phi\lesssim 0.2$ is inside the jet.  The value of
      $A_\phi\approx 0.21$ is at the disk-corona transition.  The
      value of $A_\phi\approx 0.26$ is at the equator. The plot shows
      that the poloidal current increases inside the jet, but is more
      strongly enhanced at the corona-disk interface. The current
      closes by passing through the disk in a somewhat distributed
      current sheet. }
    \label{bphiaphigrmhd}
  \end{figure}

  Figure~\ref{bphiaphigrmhd} shows the time-averaged value of $B_\phi$
  vs. the time-averaged value of $A_\phi$ at $r=10r_g$. The values
  within $A_\phi< 0.2$ correspond to the region inside the funnel that
  contains the Poynting-dominated jet.  The value of $A_\phi\approx
  0.2$ corresponds to the transition between the force-free funnel and
  the corona.  The value of $A_\phi\approx 0.21$ is at the transition
  between the corona and disk, where the increase in the poloidal
  current is largest.  Within the disk the enclosed current remains
  constant until the equator is reached and there is a strong return
  current at $A_\phi\approx 0.26$.  Notice that both hemispheres have
  been shown to demonstrate that despite the time-averaging, the
  equatorial region is still asymmetric while the jet region remains
  symmetric.  Such a plot can be used to compare GRMHD accretion
  systems to pulsar systems (see, e.g., \citealt{ckf99,mckinney2006b})
  and similar black hole force-free systems \citep{mn06}.  Despite the
  turbulence, the dependence of $B_\phi$ on $A_\phi$ remain
  surprisingly simple.  Simulations that start with a highly
  disorganized field lead to a more complicated poloidal electric
  circuit that is distributed over all angles.

  \subsection{Turbulence Leads to Simple Field Angular Velocity}

  For stationary flows, the angular velocity of the field lines
  ($\Omega_F$) is an important quantity that determines the toroidal
  field geometry and determines the electromagnetic power output of a
  jet or wind. Also, a comparison of $\Omega_F$ and the fluid angular
  velocity $\Omega$ can reveal how well-coupled the matter is to the
  field.  In principle, large scale fields can develop in the disk and
  the plasma might slip arbitrarily along the field lines, leading to
  $\Omega_F\neq \Omega$.  However, if there is strong turbulence, it
  would lead to a significant random component to the field and the
  plasma would be unable to slip as much.  Simple models of the
  accretion disk assume an arbitrary value of $\Omega_F$ in the disk,
  while here we establish which radial dependence of $\Omega_F$ is
  motivated by GRMHD simulations of turbulent accretion flows near
  rotating black holes.

  We first investigate the behavior of $\Omega_F$ and $\Omega$ as a
  function of radius in the fiducial model of \citet{mg04}.  One needs
  to choose some method to space- and time-average $\Omega_F$ in order
  to obtain its radial distribution. A poor choice would be to
  directly average $\Omega_F \equiv E_\theta/B^r$ itself because it is
  highly oscillatory and is undefined at positions where the radial
  component of the field $B^r$ momentarily vanishes.  Thus, we
  consider the ratio of space- and volume-averaged quantities to
  obtain a mean angular velocity $\langle \Omega_F\rangle \equiv
  \langle E_\theta\rangle /\langle B^r\rangle$.  For each radial shell
  this quantity is volume-averaged over a disk scale-height and
  time-averaged over approximately $8$ orbital periods at $r=10r_g$
  once the flow has reached a quasi-stationary, turbulent state. An
  alternative time-averaging is performed using the absolute value of
  each composite quantity ($|E_\theta|$ and $|B^r|$) giving
  $\langle\langle\Omega_F\rangle\rangle\equiv \langle
  |E_\theta|\rangle/\langle|B^r|\rangle$.

\begin{figure}
    \includegraphics[width=3.3in,clip]{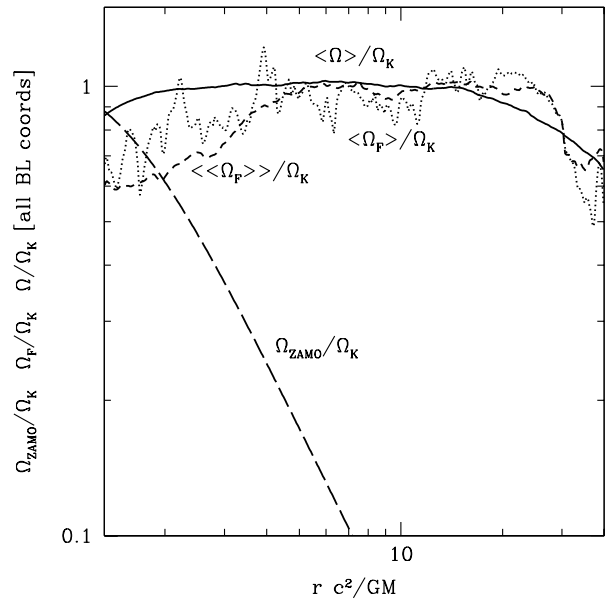}

    \caption{Angular frequencies of the plasma ($<\Omega>$, solid
      line), the field ($\langle \Omega_F\rangle$, dotted line), the
      field using a different averaging procedure
      ($\langle\langle\Omega_F\rangle\rangle$, short dashed line), and
      a ZAMO observer ($\Omega_{\rm ZAMO}$, long-dashed line) for a
      GRMHD accretion disk simulation with $a/M=0.9375$.  All
      frequencies are plotted in units of the local Keplerian angular
      frequency ($\Omega_{\rm K}$) and shown in Boyer-Lindquist
      coordinates.  Note that $<\Omega>=\Omega_{\rm ZAMO}$ on the
      horizon, as required in Boyer-Lindquist coordinates.  At large
      radii turbulence locks the field to the plasma, while at small
      radii the field rotation is locked to the spin of the black
      hole. }
    \label{grmhdomegas}
  \end{figure}

  Figure~\ref{grmhdomegas} shows the radial profiles of both forms of
  $\Omega_F$ normalized by the local Keplerian angular velocity
  $\Omega_{\rm K}$.  Both methods of averaging $\Omega_F$ lead to
  similar results.  The plot also shows the angular velocity of the
  plasma per unit Keplerian ($\Omega/\Omega_{\rm K}$) and the angular
  velocity of the zero angular momentum observer (ZAMO) per unit
  Keplerian ($\Omega_{\rm ZAMO}/\Omega_{\rm K})$. We use
  Boyer-Lindquist coordinates, with $\Omega_{\rm K} = 1/(r^{3/2}+a)$.
  Note that $\Omega_{\rm ZAMO}=\Omega_{\rm H} = a/(2r_+)$ on the
  horizon.

  As Figure~\ref{grmhdomegas} shows, $\Omega\approx \Omega_F \approx
  \Omega_{\rm K}$ over much of the disk for radii $r \gtrsim 2 r_+$.  At
  these radii, there are no large-scale fields for the plasma to slip
  along, so that the plasma and field are locked together in a turbulent
  mixture.  However, $\Omega_F$ becomes somewhat sub-Keplerian near the
  horizon as the field lines become more ordered.  Thus, while the
  plasma is forced to corotate with the black hole at the horizon, the
  field lines rotate slower with $\Omega_F\gtrsim \Omega_H/2$.  As
  discussed in \citet{mg04}, this behavior for $\Omega_F$ is consistent
  with the \citet{gam99} inflow model of the plunging region.  For
  $r\gtrsim 20r_g$, the angular velocities deviate from a simple
  behavior because the solution still depends on the initial conditions.

  As in the case of the toroidal current distribution, the results
  described here for the angular velocity of the field lines and the
  plasma are quite robust and are independent of the assumed initial
  field geometry, mass distribution, or disk thickness. However, the
  black hole spin has a dramatic qualitative effect.

  For spins from $a=-0.999$ to $a=0.999$, all the models have the plasma
  locked to the field at large radii.  However, close to the black hole
  there is a qualitative change in the results around $a/M\sim 0.4$.
  For $a/M\lesssim 0.4$ we find that $\Omega_F\approx \Omega_{\rm
    K}\approx \Omega_H$ even at the horizon because the disk dominates
  over the black hole.  However, for $a/M\gtrsim 0.4$ we find that
  $\Omega_F\approx \Omega_H/2$ near the horizon as the black hole
  dominates over the disk.  This transition at $a/M\approx 0.4$ is
  consistent with the fact that there is a qualitative change in the
  energy output of a black hole-disk-jet system at $a/M\approx 0.36$
  \citep{li2000,mckinney2005a}, when the Keplerian angular velocity of
  gas at the ISCO is equal to the angular velocity of the black hole.
  It is also consistent with the fact that there is negligible (or {\it
    negative}) electromagnetic energy extracted from the black hole for
  spins $a/M\lesssim 0.4$ for thick disks
  \citep{mg04,mckinney2005a}. Since $dP/d\theta\propto
  \Omega_F(\Omega_F-\Omega_H)$ on the horizon, we expect $dP/d\theta\sim
  0$ since $\Omega_F\sim\Omega_H$ when $a/M\lesssim 0.4$.

  \subsection{Comparison between GRMHD field and $\nu=3/4$ force-free
    field}\label{overlay}

  We have shown above that the toroidal current $I_\phi$ and the field
  angular velocity $\Omega_F$ in GRMHD simulations are well-behaved
  and so easy to model.  One expects the current distribution to be
  consistent with the nearly force-free funnel field geometry if the
  funnel field is setup directly by those currents.  Here we test the
  robustness of the association between the power-law index of the
  angle-integrated toroidal current and the polar field geometry.

  We consider a simplified problem in which neither the black hole nor
  the disk rotates.  We replace the disk with a current sheet with
  $dI_{\phi}/dr\propto r^{-5/4}$ (corresponding to $\nu=3/4$) at the
  equatorial plane and assume that the rest of the volume is in
  force-free equilibrium. In Appendix \ref{reducegrmhd}, we describe
  how to obtain force-free solutions in Schwarzschild and flat
  spacetimes for an arbitrary current sheet at the equator with no
  rotation.

\begin{figure}
    \includegraphics[width=3.3in,clip]{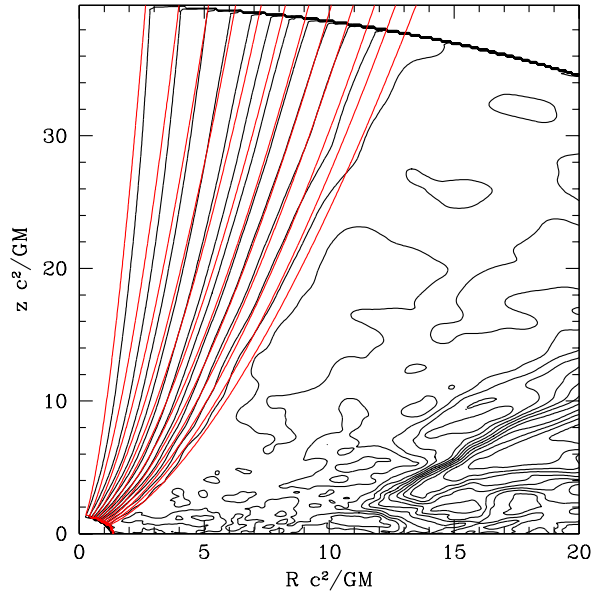}

    \caption{Overlay of the time-averaged poloidal field from the
      GRMHD numerical model (black) and the analytical $\nu=3/4$
      solution described in Appendix B.  Only the portion of the
      $\nu=3/4$ solution that overlaps the funnel region of the GRMHD
      model is shown.  Evidently, a current distribution with
      $\nu=3/4$ leads to a force-free field consistent with the funnel
      field in the GRMHD numerical model. }
    \label{overlayfields}
  \end{figure}

  Figure~\ref{overlayfields} shows the field geometry from a
  time-dependent GRMHD numerical model overlayed with the force-free
  field corresponding to $\nu=3/4$.  We see that there is a reasonably
  good agreement between the two models in the funnel region, where
  the GRMHD solution is Poynting-dominated and force-free.  The small
  differences between the models could be due to (i) residual weak
  time-dependence in the GRMHD solution, and (ii) additional
  collimation in the inner regions of the jet due to either rotation
  (which is ignored in the force-free solution plotted here) or
  coronal pressure (see \citealt{mn06}).  Also, at very small angles
  the jet is more paraboloidal.  However, the differences are small.
  In fact, the agreement between the GRMHD numerical models and the
  $\nu=3/4$ force-free model is found to be good out as far as $r\sim
  10^3r_g$ in the large scale simulations of \citet{mckinney2006c}.
  Beyond this radius the inertia of the matter becomes nonnegligible
  in the GRMHD model and the force-free approximation is no longer
  applicable.

  Although the overlay comparison shown in Figure~\ref{overlayfields}
  is impressive, we caution that the force-free field shown here is
  for the non-rotating case whereas the GRMHD solution corresponds to
  both a spinning black hole and a spinning disk.  We discuss the
  effect of rotation on force-free solutions in \citet{mn06}.

  \section{Electromagnetic Properties of the Disk}\label{emproperties}

  In \citet{mckinney2006c}, we presented a detailed study of the
  electromagnetic properties of the Poynting-dominated jet by
  providing both a qualitative description of the jet and many
  quantitative measures of the jet.  For example, radial scalings
  within the jet helped isolate the mechanism of jet formation and jet
  stability.

  In this section we first determine the radial scalings of the
  magnetic field components in the disk+corona.  We connect these
  scalings to quasi-analytic models of the accretion flow by
  \citet{gam99} that describe the disk inside the ISCO.  Given the
  simplicity of the time-averaged fields in the disk, we suggest that
  such models might be extendable to include the disk beyond the ISCO.
  Next, we compare the GRMHD simulations to simple Newtonian disk
  models that have a disk wind (such as that by \citealt{bp82}). While
  such models have a similar magnetic field scaling in the disk, the
  assumptions they make do not hold in the GRMHD simulations. Finally,
  we discuss the role of magnetic stresses near the black hole and
  compute the effective magnetic $\alpha$ viscosity parameter, which
  near the black hole rises to order unity.

  \subsection{Radial Dependence of Disk Magnetic Field}

  In this section, we focus on the radial dependence of the
  electromagnetic properties in the bulk of the disk+corona.  Thus,
  this study excludes the direct properties within the jet.  However,
  the mechanisms for disk-jet coupling may be better constrained by
  knowing the radial scalings within the disk+corona.

  We consider the space-time average of the absolute magnitude of the
  field strengths.  The volume-averaged value is found for each radial
  shell over the disk+corona, which includes only the flow that has
  $b^2/\rho_0\lesssim 1$ \citep{mg04}.  Thus the highly magnetized jet
  is explicitly excluded. For any quantity $B$ within the disk+corona
  we compute
  \begin{equation}
    \bar{B} = \frac{\int B \detg d\theta d\phi}{\int \detg d\theta
      d\phi} = \frac{\int B \sin\theta d\theta}{\int \sin\theta d\theta} ,
  \end{equation}
  which clearly preserves the radial dependence of the quantity in
  question.

\begin{figure}
    \includegraphics[width=3.3in,clip]{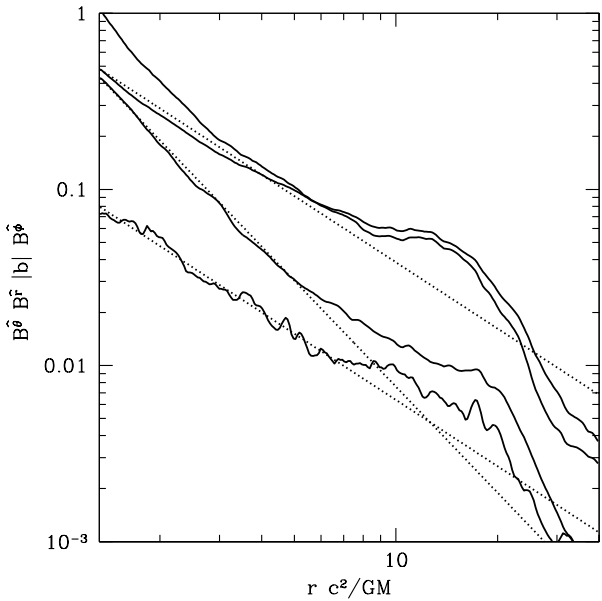}

    \caption{From above, the four solid lines correspond to
    $B^{\hat\phi}$, the comoving field strength $|b|$, $B^{\hat{r}}$,
    and $B^{\hat\theta}$ for a GRMHD accretion disk simulation with
    $a/M=0.9375$.  The magnetic field strengths roughly follow
    power-law behaviors close to the black hole.  The solution at
    $r\gtrsim 10r_g$ is still dependent on initial conditions, so this
    region is excluded from the fitting procedure.}
    \label{fields}
  \end{figure}

  As before, the time-average is computed over the turbulent period of
  accretion.  The field strengths are given in Gaussian units and
  normalized by the rest-mass density within the disk as done in
  \citet{mckinney2006c}, such that given an estimate of the density of
  the disk near the black hole or the mass accretion rate near the
  black hole one can convert to physical units.  Figure~\ref{fields}
  shows these field strengths as a function of radius; where for
  $r\gtrsim 10r_g$ the initial conditions still contribute to the
  solution and so the field there is not included in the fitting
  procedure.

  For models with any black hole spin, we find that the radial
  dependence of the comoving field and toroidal strengths roughly follow
  \begin{equation}\label{bscaling}
    \frac{|b|}{\sqrt{4\pi\rho_{0,{\rm disk}} c^2}} \approx \frac{|B^{\hat{\phi}}|}{\sqrt{4\pi\rho_{0,{\rm disk}} c^2}} \approx 0.5
    \left(\frac{r}{r_+}\right)^{-1.3} ,
  \end{equation}
  as found in \citet{mckinney2005a}.  Since the toroidal field dominates
  within the disk, the value
  of the lab-frame toroidal field ($B^{\hat{\phi}}$) is similar in
  magnitude and follows a similar dependence except inside the
  ergosphere within $r<2r_g$ where the coordinate effects of
  frame-dragging are strong for rapidly rotating black hole models.
  The angular field roughly follows
  \begin{equation}
    \frac{|B^{\hat{\theta}}|}{\sqrt{4\pi\rho_{0,{\rm disk}} c^2}} \approx 0.08
    \left(\frac{r}{r_+}\right)^{-1.3} .
  \end{equation}
  Thus the comoving, toroidal, and angular field nearly follow the $r^{-5/4}$
  power-law dependence, which is to be expected if the toroidal current
  density obeys $dI/dr\propto r^{-5/4}$.

  The radial field roughly follows
  \begin{equation}\label{brfit}
    \frac{|B^{\hat{r}}|}{\sqrt{4\pi\rho_{0,{\rm disk}} c^2}} \approx 0.4
    \left(\frac{r}{r_+}\right)^{-2.0} .
  \end{equation}
  This radial dependence is close to the monopolar scaling of
  $r^{-2}$, which is expected for a completely laminar flow where the
  mass inflow is confined to a constant $H/R$.  The monopolar radial
  field within the disk is associated with a $\nu=0$ toroidal current
  within the core of the disk. Equations (\ref{bscaling}-\ref{brfit})
  show that the toroidal field dominates the other field components
  within the disk.

  This type of solution is similar to the model proposed by
  \citet{gam99} who described a solution such that within the plunging
  region the dimensionless radial magnetic flux
  \begin{equation}
    \tilde{F}_{\theta\phi} \equiv \frac{F_{\theta\phi}}{\sqrt{-F_M}} \equiv
    \frac{r^2 B^r}{\sqrt{-2\pi r^2 \rho_0 u^r}} \left(\frac{c^{3/2}}{GM}\right)
  \end{equation}
  is a constant function of radius, where we have temporarily
  reintroduced $GM$ and $c$, $B^r$ is in Gaussian units, and $u^r$ is
  the radial 4-velocity.  Here we report that this parameter is
  nearly constant {\it throughout} the entire accretion flow out to
  $r\sim 40r_g$ with a value of
  \begin{equation}
    \tilde{F}_{\theta\phi}\approx 1.09 ,
  \end{equation}
  which is the same as found by \citet{mg04} for the plunging region
  in GRMHD simulations. The constancy of this parameter is consistent
  with a disk containing a radial field that is nearly monopolar per
  unit mass flux, as envisioned in the Gammie inflow model.

  This behavior of the accretion flow is consistent with the fact that
  the radial dependence of $b^2$ within the GRMHD plunging region
  follows the Gammie inflow solution \citep{mg04}.  As described in
  \citet{mg04}, the thin disk Gammie solution does well to model the
  GRMHD flow apart from the lack of modelling the effects of pressure
  that lead to a non-zero radial velocity across the ISCO and the lack
  of a feature in the flow near the ISCO.  One primary difference is
  that the \citet{gam99} model assumes $\Omega_F$ is also constant
  along the radial field lines, while in the turbulent disk
  $\Omega_F\sim \Omega_K$ in the outer disk.

  In summary, despite the obvious turbulence in the bulk of the disk
  and the weak-disordered field in the corona, the time-averaged
  magnetic field averaged over the disk+corona has simple radial
  scalings with a strong similarity to the model of \citet{gam99}.
  That model might be extendable to apply to the entire accretion flow
  within a disk scale height.

  \subsection{Comparison to the BP and ADAF Models}\label{bpcompare}

  The $r^{-5/4}$ power-law scaling for the electromagnetic field is of
  particular interest because it occurs rather naturally in certain
  self-similar models in the literature.  Here we check whether the
  assumptions made by these models applies to the GRMHD simulations.

  A popular model for disk winds is the \citet{bp82} (BP) self-similar
  MHD wind model in which they assumed that the \alf~speed at the
  equatorial plane scaled as the local Keplerian speed.  Coupling this
  with the additional assumption that the density scales as $\rho \sim
  r^{-3/2}$, they found that the magnetic field should scale as
  $|b|\propto r^{-5/4}$, which is consistent with the GRMHD flow as
  given by equation (\ref{bscaling}).  The BP model also assumes that
  the disk is threaded by a large-scale organized field.  Clearly this
  latter assumption is broken, as shown in figure~\ref{grmhdfield} and
  as discussed in \citet{hirose04,mckinney2005a,mckinney2006c}.

  More recently, the same scaling was obtained also in
  advection-dominated accretion flow (ADAF) models \citep{ny94,ny95a}.
  Under the assumption of self-similarity, ADAF models naturally give
  $\rho\sim r^{-3/2}$ and pressure $p\sim r^{-5/2}$.  Assuming
  equipartition between gas and magnetic pressure, one again finds
  $|b| \sim r^{-5/4}$ \citep{ny95b}.

  \begin{figure}
    \includegraphics[width=3.3in,clip]{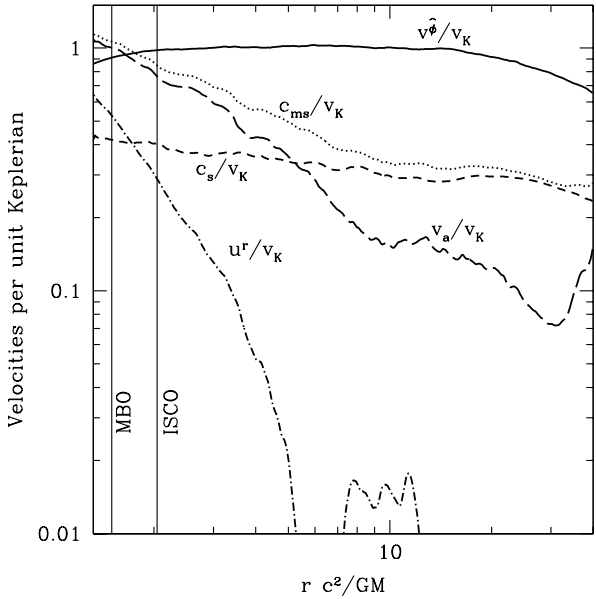}

    \caption{Radial profiles of the plasma rotational speed
      ($v^{\hat{\phi}}$, solid line), the magnetosonic speed ($c_{\rm
      ms}$, dotted line), the sound speed ($c_s$, short-dashed line),
      the \alf~speed ($v_a$, long-dashed line), and the radial
      4-velocity ($u^r$, dot - short dash line) for a GRMHD accretion
      disk simulation with $a/M=0.9375$.  All speeds are plotted in
      units of the Keplerian speed ($v_{\rm K}$).  The angular and
      Keplerian speeds are in Boyer-Lindquist coordinates, while the
      others are given in comoving coordinates. The vertical lines
      indicate the radii of the ISCO and MBO.  Note that the
      equipartition assumption, viz., $v_a\sim c_s \sim v_K$, is
      strongly violated. The solution at $r\gtrsim 20r_g$ is still
      dependent on initial conditions and is excluded from the
      analysis. }
    \label{velcsrats}
  \end{figure}

  Figure~\ref{velcsrats} shows the plasma rotational speed,
  magnetosonic speed, sound speed, \alf~speed, and in-fall radial
  4-velocity for the fiducial GRMHD numerical model. The general
  relativistic generalization of the \alf~speed is given by
  \begin{equation}
    |v_a| \equiv \frac{|b|}{\sqrt{\rho_0+u+p+b^2}}
  \end{equation}
  and the Keplerian speed is given by
  \begin{equation}
    v_K\approx \frac{r}{r^{3/2}+a} .
  \end{equation}
  Let us focus on the region interior to $r\sim 10r_g$, which has
  reached a quasi-steady-state (the region farther out is still
  sensitive to the initial conditions). Clearly, as
  Figure~\ref{velcsrats} shows, $v_a$ is not simply related to either
  $v_K$ or $c_s$.  Thus, neither the BP nor ADAF assumptions are
  satisfied.  This is perhaps not surprising since self-similar models
  assume a Newtonian gravity and so only apply far from the
  horizon. Figure~\ref{velcsrats} shows that the magnetic field
  dominates over the matter near the horizon, consistent with the
  results in \citet{dhk03} and \citet{mg04}.  This is also consistent
  with the fact that within the disk the ingoing fast magnetosonic
  surface is located at $r\approx 1.8r_g$, located between the ISCO
  and the marginally bound orbit (MBO).

  One important feature of black hole space-times is the ISCO,
  which for the simplest viscous thin disks demarcates where the fluid plunges into
  the black hole.  For our simulations of moderately thick disks
  with $H/R\approx 0.26$, we find that near the black hole within
  $r\lesssim 3r_g$ the fluid plunges into the black hole with a
  power-law radial 4-velocity,
  \begin{equation}
    u^r \approx -0.3 \left(\frac{r}{r_+}\right)^{-2} ,
  \end{equation}
  where $u^r$ is written in Boyer-Lindquist coordinates.  For the
  fiducial model with $H/R\approx 0.26$ and $a/M=0.9375$, this
  power-law plunging starts at approximately
  \begin{equation}\label{plunge}
    R_{\rm plunge} \approx R_{\rm ISCO} \left(1 +
    \left(\frac{H}{R}\right)^{0.5}\right) .
  \end{equation}
  This scaling is also consistent with the other $H/R\approx 0.05$ model we
  studied and of course is trivially consistent with thin disk theory
  for which a thin Keplerian disk has the ISCO located at $R_{\rm
  ISCO}=2.044r_g$ for $a/M=0.9375$.  Further studies can determine how general this
  fit is.  However, notice that the fluid {\it begins} to plunge
  toward the black hole at around $r\approx 6r_g$, significantly
  further out than the ISCO.  Thus there appears to be a factor of $2$
  ambiguity in identifying the location where material plunges into
  the black hole.  This GRMHD result has some not-well-defined bearing on recent
  measurements of black hole spin that used the observed spectra to
  estimate $R_{\rm ISCO}$ and so estimate $a/M$ \citep{shafee06,mcc06}.
  The ingoing slow magnetosonic and~\alf~surfaces are at $r\approx
  5r_g$ just inside where the fluid begins to plunge into the black
  hole.

  Within a scale-height and for all radii throughout the disk the mass
  accretion rate obeys
  \begin{equation}
    \dot{M}[r] \approx {\rm Const},
  \end{equation}
  which indicates that for this particular model with $a/M=0.9375$
  that mass-loss is not significant to the properties of the bulk
  flow.  For more rapidly rotating black holes the mass-loss can
  become significant \citep{hk06}, but the radial dependence in the
  disk appears non-trivial.  Since $\dot{M}[r]\propto r^2 \rho_0 u^r$,
  then one expects a roughly constant proper density and indeed
  \begin{equation}
    \frac{\rho_0}{\rho_{0,{\rm disk}}} \approx 1 \left(\frac{r}{r_+}\right)^0
  \end{equation}
  instead of $\rho_0\propto r^{-3/2}$ as one would expect in the BP or
  ADAF model.  Notice that this defines $\rho_{0,{\rm disk}}$.

  The growth in the \alf~speed is consistent with the fact that within
  $r\lesssim 10r_g$ the rest-mass density is nearly constant and the
  gas pressure is small and quickly diminishes at larger radii,
  following
  \begin{equation}
    \frac{p}{\rho_{0,{\rm disk}} c^2} \approx 0.01 \left(\frac{r}{r_+}\right)^{-2} ,
  \end{equation}
  which implies that the enthalpy is small compared to the rest-mass
  density.  An equally good fit has $p\propto r^{-1.5}$.  The comoving
  field energy is small compared to the rest-mass density, so that
  \begin{equation}
    \langle v_a \rangle \propto \langle|b|\rangle \propto r^{-5/4} ,
  \end{equation}
  for $r\lesssim 10r_g$.  Interestingly, the de-correlated average
  \begin{equation}
    \langle\langle v_a\rangle\rangle \equiv \frac{\langle |b|\rangle
    }{\sqrt{\langle \rho_0\rangle +\langle u\rangle +\langle p\rangle
    +\langle b^2\rangle }}
  \end{equation}
  follows this dependence even more strictly than the direct space-time
  average of $v_a$.  This shows that space-time correlations are mild
  between the various sources of pressure.

  We thus conclude that it is purely an accident that the field and
  the current in the GRMHD solutions scale exactly as in the BP and
  ADAF models.  Clearly the fact that $v_a\propto r^{-5/4}$ and
  $v_K\propto r^{-1/2}$ means that $v_a\sim v_K$ is not held and so
  the BP/ADAF assumptions are violated in this region close to the
  black hole. This is found to be true for many models of the disk and
  a large range of black hole spins.  It remains an open question as
  to what mechanism enforces the $\nu=3/4$ toroidal current density
  associated with the polar field.

  \subsection{Magnetic $\alpha$ viscosity parameter}

  Accretion flows without magnetic fields are often assumed to be free
  of dissipation and torques within the ISCO.  However, it has long been
  understood that magnetic fields can drastically violate this
  assumption through the action of extended fields that generate
  torques across the ISCO even without turbulent dissipation.  While
  equation (\ref{plunge}) demonstrates that the gas pressure
  scale-height is related the effective location of the ISCO, magnetic
  stresses may play some role in setting this scaling. Magnetic stresses
  may also play independent roles not at all modelled by
  a scale-height -based argument.

  An important parameter in understanding the presence of stresses
  within the ISCO is the magnitude and radial scaling of the $\alpha$
  viscosity parameter.  Early estimates of the BZ power output
  depended upon the argument that $\alpha$ was as determined in local
  shearing box calculations \citep{ga97}.  They also assumed that the
  field threading the hole was determined by a sub-equipartition
  between the field and gas pressure in the disk \citep{ga97,lop99}.
  Finally, they assumed that the electromagnetic power from the disk
  is not converted into material and thermal forms.  Based upon these
  assumptions, they concluded that BZ power output was negligible and
  certainly weaker than the electromagnetic power of the disk.  Here
  we check whether $\alpha$ behaves the same near the black hole as in
  local non-relativistic simulations.

\begin{figure}
    \includegraphics[width=3.3in,clip]{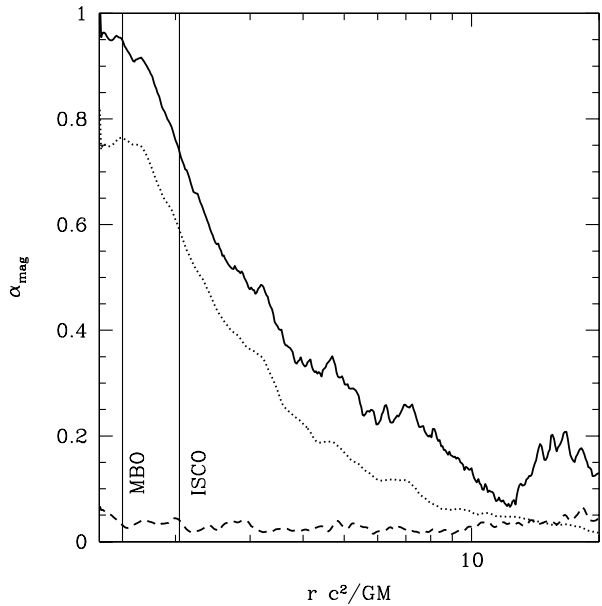}

    \caption{Radial dependence of the effective viscosity parameter
      $\alpha_{\rm mag}$ as defined in eq. (19) for a GRMHD accretion
      disk simulation with $a/M=0.9375$.  The solid line corresponds
      to the choice $P=b^2/2$, i.e., magnetic pressure alone, and the
      dotted line corresponds to $P=b^2/2+p_g$, i.e., the total
      pressure.  The dashed line shows the uncorrelated time-average
      of $-2\langle b^r\rangle \langle b_\phi\rangle /\langle
      b^2\rangle $.  All quantities have been space-time averaged over
      a period of time covering several turbulent eddy time scales.
      The vertical lines indicate the radii of the ISCO and MBO.  The
      curves show that the magnetic stress is significant inside the
      ISCO and that correlations among the field components are
      significant. } \label{alphamag}
  \end{figure}

  Local shearing box simulations of a small section of the accretion
  flow have suggested that $v_a \approx 2 \sqrt{\alpha} c_s$, where
  $\alpha\sim 0.01 - 0.1$ is the usual dimensionless viscosity in
  standard accretion disk models \citep{h95}. Global pseudo-Newtonian
  simulations have shown that $\alpha$ rises sharply inside the ISCO
  \citep{hk01}.  In our GRMHD simulations, the radial transport of
  angular momentum can be investigated by measuring the effective
  magnetic $\alpha$
  \begin{equation}
    \alpha_{\rm mag}\approx \left\langle \frac{-b^r
      b_\mu \phi^\mu}{P}\right\rangle  ,
  \end{equation}
  where $\phi^\mu=\{0,0,0,1\}$ is the $\phi$ Killing vector associated
  with the axisymmetry of the system, $b^r$ is the comoving radial
  field strength, and $b_\mu$ is the covariant comoving 4-field.  This
  form of $\alpha_{\rm mag}$ is independent of the coordinate system
  for axisymmetric space-times.  We choose $P$ to be either the total
  pressure or just the magnetic pressure $b^2/2$.  From the
  high-resolution fiducial model studied in \citet{mg04}, we compute
  the radial dependence of $\alpha_{\rm mag}$, integrated over a disk
  scale height ($H/R\approx 0.26$) and over the turbulent period of
  accretion.  The angular integration is confined to the disk, so that
  the result is not influenced by the corona above the disk.  The
  result is shown in Figure~\ref{alphamag}.  We see that $\alpha_{\rm
    mag}$ rises toward the horizon, which is consistent with the
  non-relativistic results of \citet{hk01}.  The large $\alpha_{\rm
    mag}$ is associated with a flux of angular momentum from inside the
  ISCO \citep{mg04}.

  The plot also shows the uncorrelated time-average of
  \begin{equation}
    \alpha_{\rm mag,uncorrelated} = -\left(\frac{\langle b^r\rangle
      \langle b_\phi\rangle }{\langle b^2/2\rangle }\right) ,
  \end{equation}
  where the brackets denote a time-average.  This quantity is nearly
  constant within the entire flow, showing that temporal correlations
  between the quantities are significant.

  Since the value of $\alpha$ rises near the black hole, one might
  expect that the increased stress would be associated with an
  increased energy flux from the black hole and an associated
  increased luminosity from the disk near the ISCO compared to
  standard thin disk theory \citep{krolik99,gam99}. However, for these
  GRMHD models, which correspond to radiatively inefficient accretion
  flows (RIAFs), the energy per baryon accreted is similar to that for
  a thin disk with a fixed $\alpha$ and no torques inside the ISCO
  \citep{mg04}.

  The fact that $\alpha\sim 1$ near the black hole violates the
  assumptions of \citet{ga97,lop99}.  Also, black hole spin plays a
  nontrivial role in enhancing the magnetic field threading the black
  hole horizon \citep{mckinney2005a} and as discussed in
  section~\ref{bpcompare} the field and gas pressure can be near
  equipartition near the black hole.  Finally, as discussed in
  section~\ref{powercompare}, a significant amount of the {\it disk's}
  electromagnetic power output is {\it lost} to material and thermal
  power. Thus, the BZ mechanism may still be responsible for the most
  of the electromagnetic power output from accretion systems and may
  still account for the most powerful radio sources.  These effects
  should be considered when comparing the BZ and disk electromagnetic
  powers.

  \section{Limitations}\label{limitations}

  The primary limitation of the present study is that the numerical
  models are axisymmetric.  A 3D model may show that $\nu=3/4$ is not
  generally chosen by the system.  Comparisons between 2D and 3D GRMHD
  simulations have shown reasonable qualitative and quantitative
  consistency \citep{dhk03,mg04}.  In particular, both show that the
  flow partitions into a disk, corona, disk wind, and magnetized funnel
  region.  Both give similar accretion rates of energy and angular
  momentum per unit baryon. The primarily problem with axisymmetric
  simulations is that turbulence decays after some length of time.  We
  avoid this problem by only making measurements during the turbulent
  period of accretion, so our results are unlikely to significantly
  change in 3D simulations.

  In the regime where radiation determines the energy balance in the
  disk, and so determines the disk thickness, radiative cooling should
  make little difference to these results since models with disk
  thicknesses between $H/R\sim 0.05$ to $H/R\sim 0.3$ show the same
  behavior.  However, radiative effects could also be dynamically
  important, such as through the photon bubble instability
  \citep{arons92,gam98}, and it is unknown how this would change our
  results.

  The particular field geometry accreted can significantly change the
  mass-loading of the Poynting-dominated jet.  Accreting a more random
  field leads to a larger coronal region that can extend all the way to
  the poles around the black hole, and no Poynting-dominated jet would
  form unless the black hole spin were sufficiently large that the
  toroidal magnetic pressure exceeded the ram pressure of the coronal
  material.  Preliminary models of Keplerian disks, with cooling to keep
  the thickness fixed, suggest that the accretion of an irregular field
  leads to no Poynting-jet formation for at least $a/M\lesssim 0.94$
  (McKinney \& Gammie, in prep.).  Despite the absence of the
  Poynting-jet, the toroidal current within the disk is still
  well-modelled by the $\nu=3/4$ solution.  In the tangled field case,
  the corona simply fills the region that would otherwise have been
  occupied by the Poynting-dominated jet.

  All the models studied have disks that only extend out to $r\sim
  40r_g$ with a similar circularization radius.  The results of this
  paper, such as the radial scalings, only directly apply within
  $r\lesssim 10r_g$ where general relativistic effects of the space-time
  play an important role.  The results found here may not hold far from
  the black hole.  For example, $\alpha_{\rm mag}\approx {\rm Const.}
  \sim 0.1$ is expected far from the black hole.  Also, while a more
  extended disk is not expected to affect the flow properties near the
  black hole, an extended disk may affect the Poynting-dominated jet at
  large radii.  Future studies should investigate a more extended disk
  to model systems that have a large circularization radius.

  \section{Discussion and Conclusions}\label{conclusions}

  We have shown in this paper that GRMHD numerical simulations of
  magnetized accretion flows lead to a simple angular-integrated
  toroidal current density of the form $dI_\phi/dr \propto r^{-5/4}$,
  even though the accretion disk, corona, and outflowing wind are
  highly turbulent and chaotic.  We showed that the poloidal field
  distribution in the jet is consistent with the force-free field
  solution for an $r^{-5/4}$ current distribution in a non-rotating
  equatorial disk.  The fact that we obtained good agreement even with
  a highly simplified non-rotating model suggests that
  self-collimation by hoop-stresses may not be required within the
  jet.  If hoop-stresses are ineffective, then force-balance requires
  the strongly magnetized jet to be confined by the weakly magnetized
  corona and wind (see \citealt{lyn06} for a discussion of pressure
  confinement). By measuring forces across the corona-funnel
  interface, we found that force balance primarily requires the
  coronal pressure indicating that the corona is required to
  confine/collimate the polar jet. Future work should study by what
  mechanism the confining coronal pressure generates a force-free
  field consistent with the $r^{-5/4}$ toroidal current density.

  Because of the strong turbulence, at large radii the plasma and the
  magnetic field in the disk are locked together so that both rotate
  around the black hole at roughly the local Keplerian angular
  frequency ($\Omega _K$).  For $r\lesssim 3r_g$, the behavior of the
  plasma and field qualitatively changes and depends strongly on
  whether the black hole spin is larger or smaller than $a/M\sim 0.4$.
  Near the horizon, the field angular velocity asymptotes to
  $\Omega_F\approx \Omega_H/2$ in the case of rapidly spinning holes
  with $a/M\gtrsim 0.4$, where $\Omega_H$ is the angular frequency of
  the hole, and to $\Omega_F\sim\Omega_H$ for $a/M\lesssim 0.4$.

  Although the disk is turbulent, the average magnetic field strength
  in the disk varies smoothly as $|b| \propto r^{-5/4}$, and the
  individual components of the field ($B^{\hat{r}}$, $B^{\hat\theta}$,
  and $B^{\hat\phi}$) also have simple power-law scalings.  The
  scaling of $|b|$ is similar to that assumed in the Blandford \&
  Payne (1982) and ADAF (Narayan \& Yi 1995b) models, but we showed
  that the similarity is purely accidental.  Those models assume
  Newtonian gravity and that gas and magnetic pressures scale
  similarly with radius. Gravity near the black hole is obviously not
  Newtonian, and in the simulations the magnetic pressure rises more
  rapidly then the gas pressure toward the black hole.  Thus, the
  physical reason for the particular scaling of the field with radius
  is unclear. There is some resemblance to the inflow model of
  \citet{gam99}, but the construction of a simplified 1-D model is
  left for future work.

  The dominance of the magnetic field near the horizon is associated
  with an increased magnetic stress that implies a magnetic viscosity
  parameter $\alpha\sim 1$, while $\alpha\sim 0.1$ far from the black
  hole.  The increased magnetic stress leads to enhanced angular
  momentum transport, but the energy per baryon accreted remains
  similar to that from thin disk theory (see also \citealt{mg04}).
  Prior studies that estimated the BZ power of black holes and the
  disk power output assumed an $\alpha$ based upon local shearing-box
  models that obtained $\alpha\sim 0.1$ \citep{ga97,lop99}.  These
  studies also assumed that the field strength near the black hole
  would be set by sub-equipartition arguments, but we find that the
  field near the black hole is a non-trivial function of black hole
  spin \citep{mckinney2005a}. They also assumed that the disk was
  threaded by an organized field, but we find that turbulence
  disorganizes any ordered field threading the disk. The lack of an
  organized field threading the disk is associated with a significant
  amount of electromagnetic power from the disk being converted into
  material and thermal forms. In light of all these facts, estimates
  of the BZ power (and its comparison to the disk power) should be
  reconsidered.

  The basic picture of the disk-jet coupling that emerges is that
  turbulence driven by the MRI leads to toroidal currents that in a
  force-free state would drive an organized poloidal field from the
  disk.  However, turbulence dominates the accretion flow and
  large-scale modes lead to angular momentum transport that
  continuously advects equatorial field through the disk into the hole
  and transports field at higher latitudes to larger radii.  This
  keeps the corona free of ordered fields.  The disk-corona interface
  is a site for dissipation of disk field through reconnection.  The
  energy from the dissipation drives a hot coronal wind off the disk.
  The coronal wind is thus quite different from MHD models of disk
  winds (e.g., Blandford \& Payne 1982), and is more akin to a
  thermally driven wind.  The pressure of the corona confines the
  Poynting-dominated jet, since without the corona the field gradients
  at the funnel-corona interface could not be supported.

  The geometry of the accreted magnetic field plays a crucial role in
  controlling the production of the Poynting-dominated jet.  Quite
  similar jets are obtained in GRMHD simulations that start from a
  variety of different initial configurations of the magnetic field:
  uniform vertical field, poloidal loop of magnetic field in the disk,
  multiple loops of alternating poloidal direction.  Even though the
  latter two models have zero net vertical field, nevertheless, they
  end up with a net vertical flux through the black hole and jet.  The
  mechanism by which this happens is described in the discussion of
  Model B in \citet{igumenshchev2003} and see \citet{narayan2003}.
  However, for models initialized with a mostly disorganized field,
  the Poynting-dominated jet is weaker or absent because mass
  continuously loads the polar region.  Despite the lack of a simple
  Poynting-dominated jet in such models, there is still a disk wind
  containing a disorganized field and the currents and field strengths
  within the disk still follow the same power-law dependencies.

  Although we obtain good agreement between the magnetic field
  geometry in a simple force-free current sheet model and the poloidal
  magnetic field configuration in the Poynting-dominated jet in GRMHD
  models, we note that we have not solved for the force-free {\it
  toroidal} structure of the field in the jet.  Since our force-free
  model has no rotation, $B^{\hat{\phi}}=0$ and so there is no
  Poynting flux. Also, the Lorentz factor cannot be estimated from our
  non-rotating force-free solution.  To model these quantities we must
  consider force-free models in which the disk and field are rotating
  (e.g., with a Keplerian profile).  It would be interesting to see
  how well simple rotating force-free models can reproduce the
  toroidal structure and acceleration of Poynting-dominated jets found
  in GRMHD simulations.  This is the topic of a companion paper
  (McKinney \& Narayan 2006).

  \section*{Acknowledgments}

  This research was supported by NASA-Astrophysics Theory Program
  grant NNG04GL38G and a Harvard CfA Institute for Theory and
  Computation fellowship.  We thank Serguei Komissarov, Dmitri
  Uzdensky, Vasily Beskin, Anatoly Spitkovsky, Amir Levinson, Yiannis
  Contopoulos, and the anonymous referee for useful comments and
  discussions.


  \appendix

  \section{Evolution Equations}\label{GRMHD}

  The GRMHD equations of motion are used to study magnetized accretion
  disks in the gravitational field of rapidly rotating black holes
  described by the Kerr metric using the HARM code \citep{gmt03} with
  improvements described in \citet{mckinney2006c,noble06}.  The Kerr
  metric is written in Kerr-Schild coordinates, such that the
  inner-radial computational boundary can be placed inside the horizon
  and so out of causal contact with the flow.  The Kerr metric in
  Kerr-Schild coordinates and the Jacobian transformation to
  Boyer-Lindquist coordinates are given in \citet{mg04}.

  Boyer-Lindquist coordinates are not chosen because it is difficult
  to avoid interactions between the inner-radial computational
  boundary and the jet. The coordinate singularity at the event
  horizon in Boyer-Lindquist can be avoided by placing the
  inner-radial computational boundary outside the horizon.  However,
  Poynting-dominated flows have waves that propagate outward even
  arbitrarily close to the event horizon.  Using Boyer-Lindquist
  coordinates can lead to excessive variability in the jet since the
  ingoing superfast transition is not on the computational grid, and
  then the details of the boundary condition can significantly impact
  the jet.  Numerical models of viscous flows have historically had
  related issues (see discussion in, e.g., \citealt{mg02}).

  \subsection{GRMHD Equations of Motion}

  The GRMHD notation follows \citet{mtw73}, hereafter MTW. A
  single-component MHD approximation is assumed such that particle
  number is conserved,
  \begin{equation}
    (\rho_0 u^\mu)_{;\mu} = 0 ,
  \end{equation}
  where $\rho_0$ is the rest-mass density and $u^\mu$ is the
  4-velocity. A 4-velocity with a spatial drift is introduced that is
  unique by always being related to a physical observer for any
  space-time and has well-behaved spatially interpolated values, which
  is useful for numerical schemes.  This 4-velocity is
  \begin{equation}
    \tilde{u}^i \equiv u^i -\gamma \eta^i ,
  \end{equation}
  where $\gamma=-u^\alpha \eta_\alpha$.  The additional term
  represents the spatial drift of the zero angular momentum (ZAMO)
  frame defined to have a 4-velocity of $\eta_\mu =
  \{-\alpha,0,0,0\}$, where $\alpha\equiv 1/\sqrt{-g^{tt}}$ and so
  $u^t = \gamma/\alpha$. One can show that $\gamma=(1+q^2)^{1/2}$ with
  $q^2\equiv g_{ij} \tilde{u}^i\tilde{u}^j$.

  For a magnetized plasma, the energy-momentum conservation equation
  is
  \begin{equation}\label{EOM}
    {T^{\mu\nu}}_{;\nu} = \left(T^{\mu\nu}_{\rm MA} + T^{\mu\nu}_{\rm
      EM}\right)_{;\nu} = 0.
  \end{equation}
  where $T^{\mu\nu}$ is the stress-energy tensor, which can be split
  into a matter (MA) and electromagnetic (EM) part.  In the fluid
  approximation
  \begin{equation}
    T^{\mu\nu}_{\rm MA} = (\rho_0 + u_g) u^\mu u^\nu + p_g P^{\mu\nu},
  \end{equation}
  with a relativistic ideal gas pressure $p_g=(\gamma-1) u_g$, where
  $u_g$ is the internal energy density and the projection tensor is
  $P^{\mu\nu} = g^{\mu\nu} + u^\mu u^\nu$, which projects any 4-vector
  into the comoving frame (i.e. $P^{\nu\mu} u_\mu = 0$).

  In terms of the Faraday (or electromagnetic field) tensor
  ($F^{\mu\nu}$),
  \begin{equation}\label{tmunuem}
    T^{\mu\nu}_{\rm EM} = F^{\mu\gamma}{F^{\nu}}_{\gamma}
    -\frac{1}{4}g^{\mu\nu} F^{\alpha\beta}F_{\alpha\beta},
  \end{equation}
  which is written in Heaviside-Lorentz units such that a factor of
  $4\pi$ is absorbed into the definition of $F^{\mu\nu}$, where the
  Gaussian unit value of the magnetic field is obtained by multiplying
  the Heaviside-Lorentz value by $\sqrt{4\pi}$.  The induction
  equation is given by the space components of ${\dF^{\mu\nu}}_{;\nu}
  = 0$, where $\dF^{\mu\nu} = {\frac{1}{2}}
  \epsilon^{\mu\nu\kappa\lambda} F_{\kappa\lambda}$ is the dual of the
  Faraday tensor (Maxwell tensor), and the time component gives the
  no-monopoles constraint.  Here $\mathbf{\epsilon}$ is the
  Levi-Civita tensor, where $\epsilon^{\mu\nu\lambda\delta} =
  -{\frac{1}{\detg}} [\mu\nu\lambda\delta]$ and $[\mu\nu\lambda\delta]
  $ is the completely antisymmetric symbol.  The comoving electric
  field is defined as
  \begin{equation}
    e^\mu \equiv u_\nu F^{\mu\nu} = {\frac{1}{2}}\epsilon^{\mu\nu
      k\lambda} u_\nu \dF_{\lambda k} = \eta j^\nu ,
  \end{equation}
  where $\eta$ corresponds to a scalar resistivity for a comoving
  current density $j^\mu = J_\nu P^{\nu\mu}$.  The comoving magnetic
  field is defined as
  \begin{equation}\label{bcon}
    b^\nu \equiv u_\mu \dF^{\mu\nu} = {\frac{1}{2}}\epsilon^{\mu\nu
      k\lambda} u_\mu F_{k \lambda} .
  \end{equation}
  The ideal MHD approximation, $\eta = e^\mu=0$, is assumed, and so
  the invariant $e^\mu b_\mu = 0$.  Since the Lorentz acceleration on
  a particle is $f_l^\mu=q e^\mu$, then this implies that the Lorentz
  force vanishes on a {\it particle} in the ideal MHD approximation.
  Since $e^\nu u_\nu = b^\nu u_\nu = 0$, they each have only 3
  independent components.  One can show that
  \begin{equation}
    \dF^{\mu\nu} = b^\mu u^\nu - b^\nu u^\mu ,
  \end{equation}
  and
  \begin{equation}
    F^{\mu\nu} = \epsilon^{\mu\nu\sigma\epsilon} u_\sigma b_\epsilon ,
  \end{equation}
  so that the electromagnetic part of the stress-energy tensor can be
  written as
  \begin{equation}
    T^{\mu\nu}_{\rm EM} = {\frac{b^2}{2}}(u^\mu u^\nu + P^{\mu\nu}) -
    b^\mu b^\nu .
  \end{equation}
  The other Maxwell equations,
  \begin{equation}\label{currentdensity}
    J^\mu = {F^{\mu\nu}}_{;\nu} ,
  \end{equation}
  define the current density, $J^\mu$, but are not needed in the ideal
  MHD approximation for the evolution of the matter or the magnetic
  field.

  For numerical simplicity, another set of field vectors are
  introduced, such that $B^i \equiv \dF^{it}$ and $E_i \equiv
  F_{it}/\detg$.  The two 4-vectors $e^\mu$ and $b^\mu$ and the
  3-vectors $B^i$ and $E_i$ are just different ways of writing the
  independent components of the Faraday or Maxwell tensors.  Equation
  (\ref{bcon}) implies $b^t = B^i u_i$ and $b^i = (B^i + u^i
  b^t)/u^t$. Then the no-monopoles constraint becomes
  \begin{equation}
    (\detg B^i)_{,i} = 0 ,
  \end{equation}
  and the magnetic induction equation becomes
  \begin{eqnarray}\label{induct}
    (\detg B^i)_{,t} & = & -(\detg(b^i u^j - b^j u^i ))_{,j} \nonumber\\
    & = & -(\detg(B^i v^j - B^j v^i)_{,j} \nonumber\\
    & = & -(\detg(\epsilon^{ijk}\varepsilon_k))_{,j} ,
  \end{eqnarray}
  where $v^i=u^i/u^t$, $E_i=\varepsilon_i=-\epsilon_{ijk} v^j
  B^k=-\bf{v}\times\bf{B}$ is the EMF, and $\epsilon^{ijk}$ is the
  spatial permutation tensor.  The above set of equations are those
  that are solved.  A more complete discussion of the relativistic MHD
  equations can be found in \cite{anile}.

  \subsection{Stationary, Axisymmetric Constraints}\label{stationaryaxisym}

  We now write down the Faraday tensor in terms of a vector potential
  $A_\mu$, where $F_{\mu\nu} = A_{\nu,\mu} - A_{\mu,\nu}$.  If the
  field is axisymmetric ($\del_\phi \rightarrow 0$) and stationary
  ($\del_t \rightarrow 0$), then evaluating the condition
  $\dF^{\mu\nu} F_{\mu\nu} = 0$ one finds that
  \begin{equation}
    A_{\phi,\theta} A_{t,r} - A_{t,\theta} A_{\phi,r} = 0.
  \end{equation}
  It follows that one may write
  \begin{equation}\label{omega}
    {\frac{A_{t,\theta}}{A_{\phi,\theta}}} =
    {\frac{A_{t,r}}{A_{\phi,r}}} \equiv -\Omega_F(r,\theta)
  \end{equation}
  where $\Omega_F(r,\theta)$ is an as-yet-unspecified function.  It is
  usually interpreted as the ``rotation frequency'' of the
  electromagnetic field (this is Ferraro's law of isorotation; see
  e.g. \citealt{fkr}, \S 9.7 in a nonrelativistic context).  This can
  also be written as $\Omega_F\equiv F_{tr}/F_{r\phi}\equiv
  F_{t\theta}/F_{\theta\phi}$.  As shown in \citet{mg04}, one can then
  write $F_{\mu\nu}$ in terms of the three free functions $\Omega_F,
  A_\phi$, and $B^\phi$, the toroidal magnetic field:
  \begin{equation}\label{startfmunu}
    F_{tr} = -F_{rt} = \Omega_F A_{\phi,r}
  \end{equation}
  \begin{equation}
    F_{t\theta} = -F_{\theta t} = \Omega_F A_{\phi,\theta}
  \end{equation}
  \begin{equation}
    F_{r\theta} = -F_{\theta r} = \detg B^\phi
  \end{equation}
  \begin{equation}
    F_{r\phi} = -F_{\phi r} = A_{\phi,r}
  \end{equation}
  \begin{equation}\label{endfmunu}
    F_{\theta\phi} = -F_{\phi \theta} = A_{\phi,\theta}
  \end{equation}
  with all other components zero.  Written in this form, the
  electromagnetic field automatically satisfies Maxwell's source-free
  equations.  Notice that $A_{\phi,\theta}=\detg B^r$ and
  $A_{\phi,r}=-\detg B^{\theta}$.

  \section{Reducing the GRMHD Solution to a Current Sheet}\label{reducegrmhd}

  Once we have trivialized the full GRMHD accretion disk into a
  toroidal current sheet, then we can obtain the effective stationary,
  axisymmetric poloidal magnetic field from equation
  (\ref{currentdensity}).  For an axisymmetric, stationary solution
  equation (\ref{currentdensity}) gives that
  \begin{equation}
    -\detg J^\phi = (F^{r\phi}\detg)_{,r} +
    (F^{\theta\phi}\detg)_{,\theta}
  \end{equation}
  where $F_{\theta\phi} = A_{\phi,\theta} = \detg B^r$, $F_{r\phi} =
  A_{\phi,r} = -\detg B^\theta$, $F^{r\phi} = g^{\mu r} g^{\nu \phi}
  F_{\mu\nu}$. The equation in Boyer-Lindquist becomes
  \begin{eqnarray}
    \detg J^\phi & = &\nonumber\\
    & - & (|g| g^{rr} B^\theta(\Omega_F g^{t\phi}-g^{\phi\phi}))_{,r} \nonumber\\
    & + & (|g| g^{\theta\theta} B^r (\Omega_F
    g^{t\phi}-g^{\phi\phi}))_{,\theta} ,
  \end{eqnarray}
  or in terms of the vector potential and $\Omega_F$,
  \begin{eqnarray}
    \detg J^\phi & = &\nonumber\\
    & + & (\detg g^{rr} A_{\phi,r}(\Omega_F g^{t\phi}-g^{\phi\phi}))_{,r} \nonumber\\
    & + & (\detg g^{\theta\theta} A_{\phi,\theta} (\Omega_F
    g^{t\phi}-g^{\phi\phi}))_{,\theta} .
  \end{eqnarray}

  In either Boyer-Lindquist or Kerr-Schild coordinates, the linear
  partial differential equation (PDE) for $a/M=0$ is
  \begin{equation}\label{GRADSHAF1}
    -r^2\sin\theta J^\phi  = \left(\frac{r-2M}{r \sin\theta} A_{\phi,r}
    \right)_{,r} + \left(\frac{1}{r^2 \sin\theta} A_{\phi,\theta}
    \right)_{,\theta} ,
  \end{equation}
  and for a current sheet,
  \begin{equation}\label{GRADSHAF2}
    -r^2 K^\phi\delta(\theta-\pi/2)  = \left(\frac{r-2M}{r \sin\theta}
    A_{\phi,r} \right)_{,r} + \left(\frac{1}{r^2 \sin\theta}
    A_{\phi,\theta} \right)_{,\theta} .
  \end{equation}
  This equation is an elliptic partial differential equation for
  $r>2M$ and is hyperbolic for $r<2M$.  The quantity $K^\phi$ is the
  surface current density on the equatorial current sheet.

  \subsection{Vacuum Solutions in Schwarzschild space-time}

  Equation (\ref{GRADSHAF2}) can be solved to find the complementary
  solution by setting the quantity on the right to $0$ everywhere except
  on the current sheet at $\theta=\pi/2$ and assuming that the solution
  is separable such that
  \begin{equation}
    A_\phi(r,\theta)=R(r)\Theta(\theta) .
  \end{equation}
  Since the PDE is second order, there are in general two free
  functions.  The radial ordinary differential equation (ODE) can be
  written in closed form in terms of generalized hypergeometric
  function or solved numerically for $M\neq 0$.  In the limit of
  $x\rightarrow \infty$, one finds a complementary function of
  \begin{equation}\label{Rr}
    R(r) = C_0 r^{-l} + C_1 r^{l+1} .
  \end{equation}
  where $C_0$ and $C_1$ are arbitrary constants and $l\ge -1/2$. These
  are the standard spherical radial eigenfunctions.  The angular
  complementary function can be written in terms of generalized
  hypergeometric functions or alternatively written in terms of the
  associated Legendre functions of the first ($P$) and second ($Q$)
  kind, where
  \begin{equation}\label{leg}
    \Theta(\theta) = D_0 |\sin\theta| P_l^1(\cos\theta) +
    D_1|\sin\theta| Q_l^1(\cos\theta) ,
  \end{equation}
  where $D_0$ and $D_1$ are different arbitrary constants and where
  $l\ge -1/2$ are linearly independent (i.e. $P^\mu_{-l-1} =
  P^\mu_l$). This form is a mixture of two hypergeometric functions
  for each $l$ (see, e.g., chpt. 8 in \citealt{as72} and in
  \citealt{gr94}).  These functions are just one of the vector
  spherical harmonics.

  \subsection{Constraints}

  This general solution is a sum of any coefficients for all $l$, but
  has no constraints to avoid divergences (e.g. monopoles or
  divergences in the physical field strength) on the coordinate
  singularities.  The only constraint required is that $F^{\mu\nu}
  F_{\mu\nu}$ remains finite, where in Boyer-Lindquist coordinates
  with $a/M=0$ and $B^\phi=0$,
  \begin{equation}\label{fsqconstraint}
    F^{\mu\nu} F_{\mu\nu}  =  2( (F^{r\phi})^2 g_{rr} g_{\phi\phi} +
    (F^{\theta\phi})^2 g_{\theta\theta} g_{\phi\phi}) ,
  \end{equation}
  and so to avoid divergences at the polar axes one requires
  $A_\phi\propto \theta^{2+n}+{\rm Const.}$ for $n\ge 0$. There is no
  requirement on the horizon and no constraint on divergence need be
  placed at $r=0$ since it is a physical singularity.

  The solution given by equation (\ref{leg}) with $\sin(\theta)$
  multiplied by the associated Legendre functions of the first kind
  gives $A_\phi\propto \theta^2+O(\theta^4)$ for any $l$, while the
  term with the second kind gives $A_\phi\propto {\rm
    Const.}+O(\theta^2)$ for all $l$.  Hence, the only valid solutions
  are of the first kind with any radial solution, except for angular
  solutions combined with a radial function that does not depend on
  radius.  The single nontrivial example of this exception is the
  monopole solution with $l=-1$ and $C_0=D_0=0$ giving
  $A_\phi=-\cos\theta$.  The remaining solutions require the first
  kind with $D_1=0$.  The monopole solution is only a result of the
  limit to the open set around $l=-1$ for $C_1\neq 0$ or $l=0$ for
  $C_0\neq 0$ for the associated Legendre function of the first kind.
  This issue regarding the monopole solution can be considered a
  pathology of using the Legendre functions that is not manifested in
  the hypergeometric form of the solution, where $l=0$ naturally
  generates the monopole and paraboloidal type solutions.

  \subsection{Currents}\label{currents}

  The current density given by equation (\ref{currentdensity}) can be
  integrated to determine the net toroidal current enclosed within the
  volume between $r_0$ and $r$, as given by equation~\ref{didr}.  For a
  model with a current sheet located at $\theta=\pi/2$ with
  \begin{equation}
    J^\phi = K^\phi \delta(\theta-\pi/2) ,
  \end{equation}
  where $K^\phi$ is the surface current density, then
  \begin{equation}
    \frac{dI_{\phi}}{dr} = \detg[\theta=\pi/2] K^\phi .
  \end{equation}

  For a general power-law toroidal current per unit radius with
  \begin{equation}\label{diencdr}
    \frac{dI_{\phi}}{dr} = \frac{C}{r^{n+1}} =
    \frac{C}{r^{2-\nu}} ,
  \end{equation}
  the enclosed current (for $n\neq 0$) is
  \begin{equation}\label{iencvsr}
    I_{\phi}(r) \equiv I_{\phi}(r_0) +
    \left(I_{\phi}(r_2)-I_{\phi}(r_0)\right)\left(\frac{{\frac{1}{r_0^n}-\frac{1}{r^n}}}{{\frac{1}{r_0^n}-\frac{1}{r_2^n}}}\right)
    ,
  \end{equation}
  where $r_2$ denotes some radius for which $r_0<r<r_2$ and for which
  the enclosed toroidal current $I_{\phi}(r_2)$ is known. If
  $a/M=0$ and in spherical polar coordinates, then
  $\detg[\theta=\pi/2]=r^2$, so that if the toroidal surface current
  density is
  \begin{equation}\label{kphi}
    K^\phi = \frac{C}{r^{n+3}} = \frac{C}{r^{4-\nu}} .
  \end{equation}

  \subsection{Split-Monopole Field}

  The split-monopole solution for $a/M=0$ is a solution where the
  current sheet at $\theta=\pi/2$ has
  \begin{equation}\label{jphimono}
    J^\phi = \frac{C}{r^4} \delta(\theta-\pi/2) ,
  \end{equation}
  so that
  \begin{equation}
    \frac{dI_{\phi}}{dr} = \frac{C}{r^2} ,
  \end{equation}
  which gives a total enclosed current of
  \begin{equation}
    I_{\phi} = \frac{C}{\frac{1}{r_0}-\frac{1}{r}} ,
  \end{equation}
  where $r_0<r$ are the inner- and outer- radial positions.  The
  split-monopole has $n = 1$ and $\nu=0$.

  The linearly independent particular solution for the split-monopole
  vector potential (for any $M$) is
  \begin{equation}
    \label{aphisplit} A_\mu^{\rm (split)} =
    \begin{cases}
      -C\cos{\theta}\phi_\mu & \text{$\theta < \pi/2$} \\
      +C\cos{\theta}\phi_\mu & \text{$\theta > \pi/2$} ,
    \end{cases}
  \end{equation}
  where $\phi^\mu=\{0,0,0,1\}$ is the $\phi$ Killing vector.  The
  above gives that $B^\theta=B^\phi=0$ and that
  \begin{equation}
    \label{brsplit} B^r =
    \begin{cases}
      +\frac{C}{r^2} & \text{$\theta < \pi/2$} \\
      -\frac{C}{r^2} & \text{$\theta > \pi/2$} ,
    \end{cases}
  \end{equation}

  This solution has field lines that have an opening angle following
  \begin{equation}
    \theta_j\propto r^{0} .
  \end{equation}

  \subsection{BZ Paraboloidal Field}

  The paraboloidal field solution for $a/M=0$ is a solution where the
  current sheet at $\theta=\pi/2$ has
  \begin{equation}
    J^\phi = \frac{C}{r^3} \delta(\theta-\pi/2) ,
  \end{equation}
  so that
  \begin{equation}
    \frac{dI_{\phi}}{dr} = \frac{C}{r} .
  \end{equation}
  Notice that the total enclosed toroidal current would diverge for a
  disk of infinite radial extent, i.e.,
  \begin{equation}
    I_{\phi} = C\log\left(\frac{r}{r_0}\right) ,
  \end{equation}
  where $r_0<r$ are the inner- and outer- radial positions.  The
  paraboloidal solution has $n = 0$ and $\nu=1$.

  The linearly independent particular solution for the
  BZ paraboloidal field vector potential is
  \begin{equation}
    \label{aphipara} A_\mu^{\rm (para)} =
    \begin{cases}
      +g(r,\theta)\phi_\mu & \text{$\theta < \pi/2$} \\
      +g(r,\pi-\theta) \phi_\mu & \text{$\theta > \pi/2$} ,
    \end{cases}
  \end{equation}
  where $g(r,\theta)\equiv \frac{C}{2}[r f_- + 2M f_+(1-{\rm ln}f_+)]$,
  $f_+=1+cos\theta$, $f_-=1-cos\theta$, and $M$ is the mass of the
  black hole.  Thus, $B^\phi=0$ and
  \begin{equation}
    \label{brpara} B^r =
    \begin{cases}
      +\frac{C(r+2M{\rm ln}f_+)}{2 r^2} & \text{$\theta < \pi/2$} \\
      -\frac{C(r+2M{\rm ln}f_-)}{2 r^2} & \text{$\theta > \pi/2$} ,
    \end{cases}
  \end{equation}
  and
  \begin{equation}
    \label{bhpara} B^\theta =
    \begin{cases}
      +\frac{C \tan(\theta/2)}{2 r^2} & \text{$\theta < \pi/2$} \\
      +\frac{C \cot(\theta/2)}{2 r^2} & \text{$\theta > \pi/2$} ,
    \end{cases}
  \end{equation}

  This solution has field lines that have an opening angle that
  at large radii approximately obeys
  \begin{equation}
    \theta_j\propto r^{-0.5} .
  \end{equation}

  \subsection{Constructing Current Sheets by Splicing Source-Free Solutions}\label{CONSTRUCT}

  While in general it is difficult to find $A_\phi$ for arbitrary
  source functions, one can construct source functions corresponding
  to equatorial current sheets by splicing a single equatorially
  asymmetric vector potentials $G(r,\theta)$.  Then the general
  solution with a current sheet is
  \begin{equation}
    A_\phi(\theta) = G(\theta) (1-H\!(\theta-\pi/2)) + G(\pi-\theta)
    H\!(\theta-\pi/2) ,
  \end{equation}
  for either $G$ in the range of $0\le\theta\le\pi/2$ or
  $\pi/2\le\theta\le\pi$, where $H$ is the Heaviside function.

  First the simple $a/M=M=0$ equations are considered.  To construct a
  radial power-law current density of the form given by equation
  (\ref{kphi}) so that the source function is $\detg J^\phi \propto
  r^{\nu-2}$, the radial complementary or particular functions must
  satisfy
  \begin{equation}
    R(r) \propto r^{\nu} ,
  \end{equation}
  at large radii ($r\gg M$) as demonstrated by plugging this form of
  $A_\phi$ into equation (\ref{GRADSHAF1}).  This means that one must
  choose either
  \begin{equation}
    l = -\nu
  \end{equation}
  with $C_0\neq 0$ and $C_1=0$ or one must choose
  \begin{equation}
    l = \nu-1
  \end{equation}
  with $C_1\neq 0$ and $C_0=0$, where $l\ge -1/2$.  Hence, for
  $\nu>1/2$, only the second choice leads to a real solution.  For
  $\nu < 1/2$, only the first choice leads to a real solution.  For
  $\nu=1/2$ both give $l=-1/2$.  Thus, for fixed $\nu$, there is a
  unique $l$ that gives a single allowed $R(r)\propto r^{\nu}$ and
  $\Theta(\theta)$.

  For example, for $\nu=1$ one has that $l=0$, and then the most
  general solution for $M=0$ is
  \begin{equation}
    A_\phi = (c_0 + c_1 r) (d_1 + d_2 \cos{\theta}) ,
  \end{equation}
  which after forcing $A_\phi\propto \theta^{3-\nu}+{\rm Const.}$ for
  $\nu\le 1$ near $\theta=0$, one has that
  \begin{equation}
    A_\phi = (c_0 + c_1 r) (\cos{\theta}-1) ,
  \end{equation}
  which for $c_0=0$ gives paraboloidal solution in the upper
  hemisphere, $c_1=0$ gives the monopole solution, while combinations
  give a mixture of monopole and paraboloidal solutions.  Another
  example is $C_0=D_1=0$ and $l=1/2$ giving a decollimating field
  geometry.

  The GRMHD numerical solutions are associated with models with
  $\nu=3/4$, for which one must choose $l=-1/4$ and $C_0=0$.  For the
  solution to satisfy regularity on the axis near $\theta=0$, one must
  set
  \begin{equation}
    D_0\neq 0~~~{\rm and}~~~D_1=0 .
  \end{equation}
  This solution is quite similar to the paraboloidal solution, but
  slightly less collimated, as expected.  This solution has field
  lines that have an opening angle that approximately follows
  \begin{equation}
    \theta_j\propto r^{-0.375} .
  \end{equation}

  This force-free model is also used in \citet{mn06} to find
  force-free solutions with arbitrary $M$ and $a/M$ using a general
  relativistic force-free electrodynamic code
  \citep{mckinney2006b,nmf06}.



    \label{lastpage}

  \end{document}